\documentclass[aps,prl,onecolumn,showpacs,superscriptaddress, preprint]{revtex4-2}

\usepackage{graphicx}   
\usepackage{amsmath}    
\usepackage{amssymb}    
\usepackage{hyperref}   
\usepackage{bm}         
\usepackage{siunitx}    
\usepackage{setspace}
\begin{document}

\title{Quantitative 3D imaging of highly distorted micro-crystals\\using Bragg ptychography}

\author{Peng Li}
\thanks{Corresponding author: peng.li@ustc.edu.cn}
\affiliation{National Synchrotron Radiation Laboratory, University of Science and Technology of China, Hefei, China}

\author{David Yang}
\thanks{Current address: Condensed Matter Physics and Materials Science Department, Brookhaven National Laboratory, Upton, NY 11973, USA}
\affiliation{Department of Engineering Science, University of Oxford, Oxford OX1 3PJ, UK}

\author{Christoph Rau}
\affiliation{Diamond Light Source, Harwell Science and Innovation Campus, Didcot, UK}

\author{Marc Allain}
\affiliation{Aix-Marseille Univ, CNRS, Centrale Med, Institut Fresnel, Marseille, France}

\author{Felix Hofmann}
\affiliation{Department of Engineering Science, University of Oxford, Oxford OX1 3PJ, UK}

\author{Virginie Chamard}
\affiliation{Aix-Marseille Univ, CNRS, Centrale Med, Institut Fresnel, Marseille, France}

\date{\today}

\begin{abstract}
Bragg coherent diffraction imaging (BCDI) fails to reliably retrieve phases in micro-crystals exhibiting strong strain inhomogeneities, which restricts its applicability. Here we show that three-dimensional Bragg ptychography (3DBP) overcomes this limitation by enabling stable inversion for large lattice distortions. Using a combination of experimental measurements and numerical tests, we compare the performance limits of the two approaches and demonstrate that 3DBP tolerates lattice distortions more than six times larger than BCDI. We also establish the sensitivity of both methods on a weakly distorted crystal, for which 3DBP yields smoother amplitude and phase fields with reduced short-length-scale artifacts. 3DBP thus provides a reliable route for imaging micro-crystals with large lattice distortions, expanding the scope of coherent X-ray Bragg microscopy to strongly deformed systems.
\end{abstract}

\maketitle



\textit{Introduction} - Crystalline particles with size exceeding a few hundreds nanometres plays a central role in many rapidly evolving scientific fields. In fundamental research, their relevance spans plasmonic phenomena governed by particle surface structure \cite{Ostovar_sciadv_2024} to bio-production of sub-micrometric magnetic biominerals through tightly regulated biochemical pathways \cite{uebe_microbiol_2016, Pei_communications_2025}. Their technological importance is equally broad. Phase separation in lithium iron phosphate platelets limits battery performance \cite{Yu_nanolett_2015, Hughes_Materials_2022}. 
Grain scale effects play a central role in determining the integrity of modern engineering alloys. For example, hydrogen embrittlement is a key challenge for materials used in hydrogen fuel systems, arising from the interaction of hydrogen with crystal defects \cite{Yang_advmater_2025}. High-temperature aerospace alloys rely on engineered microstructures for strength and thermal stability \cite{Perrut_CRPHys_2018}. 
These examples underline how the relationship between structure, properties and performance require the precise knowledge of the lattice distortions induced by interfaces and defects, within micro-crystals and micro-crystalline aggregates. A wide range of crystal morphologies is concerned -- including isotropic particles, platelets, wires, and rods -- with typical sizes spanning from a few tens to about one thousand nanometers. For small nanoparticles (typically below a few 10's of nm), recent advances in electron microscopy have enabled three-dimensional (3D) imaging by tomography \cite{Jo_NatCommun_2022} and operando two-dimensional characterization by 4D-STEM \cite{Perich_NanoLett_2025}. However, the investigation of thicker systems relies on two-dimensional approaches that require additional sample thinning. To address this issue, synchrotron-based X-ray microscopy techniques are continuously being developed and refined. This is particularly exemplified by recent dark-field X-ray microscopy results \cite{shukla_2025}. But efforts are needed to further improve spatial resolution and sensitivity to strain and lattice distortions.

In this context, lensless strategy approaches, commonly referred to as Bragg coherent diffraction imaging (BCDI) \cite{pfeifer06, Hofmann_multi_Bragg_2017, Cha_energy_2015} are excellent alternatives. They generate quantitative highly-resolved 3D images of crystalline particles from the numerical inversion of diffraction intensity patterns acquired in the Fraunhofer regime, in the vicinity of a chosen Bragg diffraction peak \cite{pfeifer06}. To solve the phase problem, \textit{i.e.}, to ensure retrieval of the complex-valued diffracted field from the measurement of its squared amplitude only, the data need to be oversampled by at least twice the Nyquist criterion \cite{pfeifer06}. The diffracted field, amplitude and phase, is further numerically back propagated to the sample plane to produce the sample image. These methods have been successfully deployed to track particle surface evolution during catalysis or electrochemical processes \cite{Simonne_catalysis_2024, Atlan_electrochem_BCDI_2023}, dislocation formation under external loading \cite{Yehya_nanoindentBCDI_2024}, or dissolution mechanisms in mineral crystallites \cite{Clark_dissolution_2015}. However, BCDI remains challenging, as full \textit{operando} studies require imaging across a broad range of strain fields, from weakly to highly inhomogeneous, often exceeding the empirically estimated lattice-displacement threshold of approximately 0.5–1 times the unstrained lattice parameter \cite{Cha_2010, newton_prb_10, Huang_prb_2011, Pavlov_SR_2017, Zhao_modulator_2025}. This constitutes a major limitation of the method \cite{Suzana_calcite_2024}.

To overcome this, 3D Bragg ptychography (3DBP) has been proposed as a means to image crystalline materials even in the presence of highly inhomogeneous strain fields, without compromising the performances already achieved by BCDI \cite{mastropietro_NatMat_17, Hill2018, li21}. It makes use of a finite, time-invariant structured illumination beam that is scanned across the sample to introduce spatial diversity into the data set. \cite{godard_nat_11, Hru16}. To solve the phase problem, sufficient illumination overlap must be ensured between successive regions on the sample \cite{guizar_sicairos_2026}. The numerical inversion combines the full set of diffraction patterns to retrieve both a quantitative image of the extended crystalline sample and the illumination field \cite{li21}. While 3DBP could be a powerful tool for revealing the development of crystalline distortions -- owing to its high sensitivity to weak strains and lattice tilts, as well as its robustness to large lattice distortions -- the use of 3DBP to image isolated crystalline particles has not yet been demonstrated. In this letter, we present reconstructions and analysis of 3D crystalline maps produced on isolated particles with 3DBP and compare the performance of this method to state-of-the-art BCDI. 

\begin{figure}
    \centering
    \includegraphics[width=0.5\linewidth]{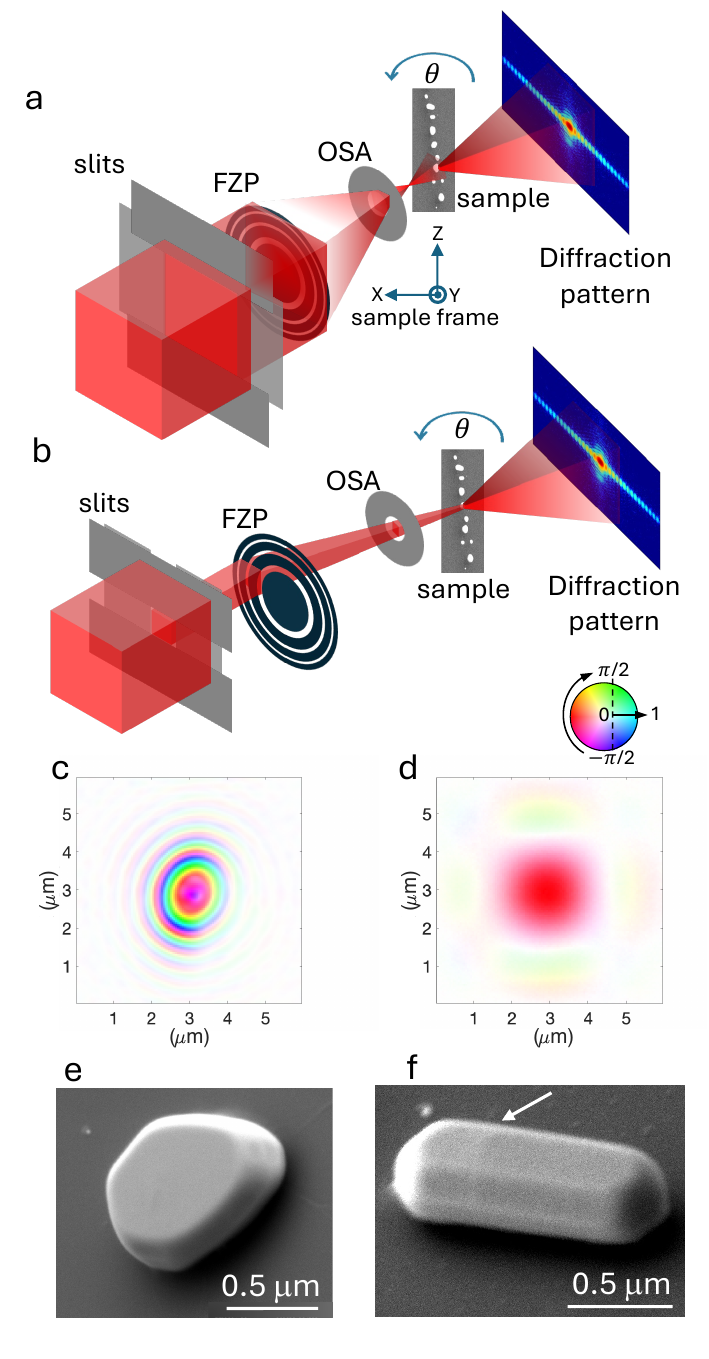}
    \caption{\textbf{Experimental setup}. (a) 3DBP set-up including a fully illuminated Fresnel zone plate (FZP), an order sorting aperture (OSA), the sample installed vertically on a piezo-stage and a rotation stage ($\theta$). The position of the sample is shifted downstream the focal plane to increase the illumination spot onto the sample. The diffracted intensity pattern is measured with a 2D detector placed in the far field. The same configuration is used to produced the curved-beam BCDI data set presented in Supplemental Materials, section S2 \cite{Supplemental_Materials}. (b) Same set-up optimized for the plane-wave BCDI acquisition. The aperture of the FZP is strongly reduced to enlarge the spot size at the focal plane, producing a locally flat illumination over an area much larger than the particle size. (c, d) Retrieved wave-fields at the sample position for configurations (a) and (b), respectively. (e, f) Scanning electron microscopy image of the weakly and highly distorted particles, respectively. In (f) the white arrow points towards a faint contrast visible at the surface of the particle.}
    \label{fig:1}
\end{figure}

\textit{Experiment} - The experiment was performed on Au particles at the I13-1 beamline at Diamond Light Source, UK. The sample preparation, which is described in detail in Appendix A
, leads to faceted micro-crystals with size ranging from $\approx$ 100~nm to a few $\mu$m and [111] crystal planes parallel to the substrate. Details of the experimental geometry are given in Appendix B. The 11.8~keV incident X-ray photon beam was focused by a Fresnel zone plate producing a focal spot of about 250~nm, smaller than the particle size (see Fig.~\ref{fig:1}(a) and \ref{fig:1}(b)). For 3DBP acquisitions (Fig.~\ref{fig:1}(a)), the sample position was set a few millimeters downstream the focal plane, to reduce the number of positions and relax the scanning accuracy constraint imposed by the ptychography scan. This allowed producing a 2~$\mu$m divergent wave-field (see Fig.~\ref{fig:1}(c) and Supplemental Materials, section S1 \cite{Supplemental_Materials}). For BCDI, which requires a plane wave illumination on the particle, the focal spot produced by the lens was enlarged by reducing the lens aperture (Fig.~\ref{fig:1}(b)). The resulting wave-field is shown in Fig.~\ref{fig:1}(d). It presents a $2\times 2~\mu m^2$ central region featured with rather constant amplitude and phase. For both BCDI and 3DBP, the sample was mounted vertically and further rotated to the Bragg angle of the symmetric (111) reflection. An area detector was positioned at twice the Bragg angle in the far field. For the BCDI measurements, an angular rocking scan was carried out, rotating the sample about the vertical axis and recording the diffraction pattern at each angular position. For 3DBP, in addition to the angular scan designed for BCDI, a raster grid scan was performed at each angle, scanning the sample parallel to the Au substrate plane. Exposure times were tuned so that the total photon counts were equivalent across the 3DBP and BCDI datasets, ensuring fair comparison of their respective performance. Further details regarding data acquisition are given in Appendix B. For each imaging modality, two crystalline particles were investigated, corresponding respectively to a weakly or highly distorted crystal. From scanning electron microscopy observations, shown in Fig.~\ref{fig:1}(e) and \ref{fig:1}(f), the weakly distorted crystal, although well-faceted, is laterally rather isotropic (Fig.~\ref{fig:1}(e)). In contrast to this, the highly distorted crystal is elongated along one dimension and presents a faint contrast, highlighted by a white arrow in Fig.~\ref{fig:1}(f)). It likely indicates a boundary between two sub-crystals, probably tilted with respect to each other. Each produced data set was inverted with numerical iterative routines described in Appendix C. 

\begin{figure}
    \centering
    \includegraphics[width=0.9\linewidth]{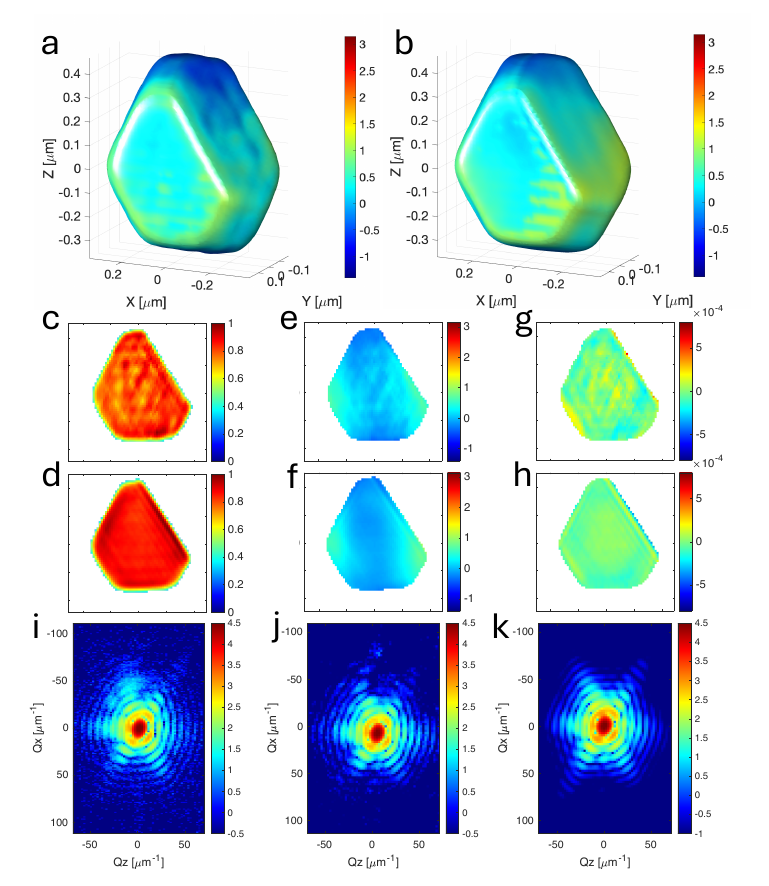}
    \caption{\textbf{3D imaging of a weakly distorted particle}. (a, b) 3D reconstructions of the weakly strained crystal showing the isosurface of the amplitude, colored with the reconstructed
phase, for BCDI and 3DBP respectively. (c, d), (e, f) and (g, h) 2D cross sections of the retrieved amplitudes, phase (in radians) and strain, respectively. From (c) to (h), the top and bottom rows correspond to BCDI and 3DBP reconstruction, respectively. (i) Diffraction pattern measurement from the plane wave BCDI data set. (j) Simulation of the BCDI diffraction pattern calculated from the plane wave BCDI reconstruction and (k) from the 3DBP reconstruction.}
    \label{fig:4}
\end{figure}

The suitability of 3DBP for imaging isolated particles is first assessed on the weakly distorted crystal. The BCDI and 3DBP reconstruction results are presented in Fig.~\ref{fig:4} plotted in the sample $(\mathbf{X}, \mathbf{Y}, \mathbf{Z})$ frame with $\mathbf{Y}$ aligned along the (111) Bragg reflection. The 3D retrieved amplitude and phase maps are respectively associated to the effective electron density of the sample crystalline part and its internal displacement field \cite{pfeifer06}. More precisely, the phase corresponds to the projection of the displacement field onto the chosen Bragg vector (\textit{i.e.}, the (111) Bragg reflection in the present study). As expected, BCDI successfully retrieves the object, exhibiting a reconstructed morphology in excellent agreement with electron microscopy observations (Fig.~\ref{fig:1}(e)). Its phase spans a range of $0.15 \times 2\pi$ (Fig.~\ref{fig:4}(e)), while the total strain range, calculated from the phase derivative \cite{Pateras15}, is on the order of $6 \times 10^{-4}$ (Fig.~\ref{fig:4}(g)). Good agreement is also found between the BCDI measurement and the coherent diffraction pattern calculated from the BCDI-retrieved object (Fig.~\ref{fig:4}(i) and \ref{fig:4}(j), respectively), both plotted in the space conjugated to the $(\mathbf{X}, \mathbf{Y}, \mathbf{Z})$ frame and noted $(\mathbf{Q_X}, \mathbf{Q_Y}, \mathbf{Q_Z})$.
The 3DBP reconstruction results, shown in Figs.~\ref{fig:4}(b), \ref{fig:4}(d), \ref{fig:4}(f) and \ref{fig:4}(h), demonstrate that Bragg ptychography also reliably retrieves the 3D complex-valued map of the isolated crystal. Moreover, the 3DBP reconstruction exhibits smoother amplitude and phase fields than BCDI reconstruction, which displays short-length-scale fluctuations that translate into additional strain components (see the yellow features in Fig. \ref{fig:4}(c), also visible in Figs. \ref{fig:4}(e) and \ref{fig:4}(g)). This possibly results from differences in the way experimental uncertainties propagate to the result in BCDI and 3DBP processes. While detection imperfections (\textbf{e.g.} Poisson shot noise) are fully retrieved in BCDI and assigned to the sample, they are smoothed out in the 3DBP process, because 3DBP relies on inverting data from the same illuminated region but associated with different realizations of the shot noise. The BCDI diffraction pattern, simulated by using the particle retrieved with the 3DBP data set (Fig.~\ref{fig:4}(k)) is in excellent agreement with the BCDI measurement shown in Fig.~\ref{fig:4}(i). Another indication of the reconstruction quality for the 3DBP method is the excellent agreement between the probe retrieved from the 3DBP data set and that from the forward ptychography as shown in the Supplemental Materials, section S1 \cite{Supplemental_Materials}. As probe and object are linked in 3DBP, it also validates the quality of the retrieved object.

\begin{figure}
    \centering
    \includegraphics[width=1.0\linewidth]{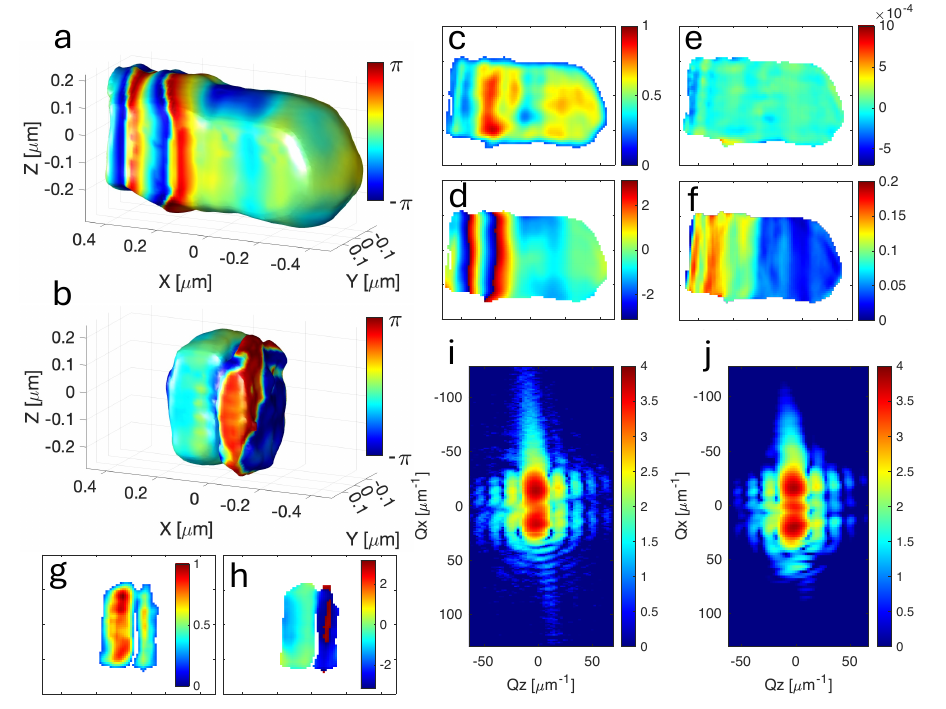} 
    \caption{\textbf{Quantitative 3D imaging of a highly distorted micro-crystal.} (a, b) 3D reconstructions of the highly distorted crystal showing the iso-surface of the amplitude, colored with the reconstructed phase, obtained from 3DBP and BCDI, respectively. (c, d, e, f) 2D cross-sections of the amplitude, phase, strain and lattice tilt for the 3DBP reconstruction. (g, h) 2D cross-section of the amplitude and phase for the BCDI reconstruction. (i) Diffraction pattern measurement from the plane wave BCDI data set and (j) BCDI diffraction pattern simulated by using the particle retrieved with the 3DBP data. Phase maps are in radians. In (i, j) intensity unit is photon counts on logarithmic scale.}
    \label{fig:2}
\end{figure}

The case of the highly distorted crystal is presented in Fig.~\ref{fig:2}. The 3D reconstruction obtained from the 3DBP data set evidences the presence of an inhomogeneous phase distribution within the $\approx 1~\mu$m long crystal (Fig.~\ref{fig:2}(a) and \ref{fig:2}(d)): about half of the crystal is characterized by a rather constant phase while the other part exhibits a phase distribution, which linearly increases along the $\mathbf{X}$ axis. This phase gradient of 0.7 rad per retrieved pixels (or 0.04 rad/nm) develops over 0.4 $\mu$m, exploring a phase range of about $2.4 \times 2 \pi$. It corresponds to a tilt of the lattice planes about the $\mathbf{Z}$ axis. Plots of the retrieved amplitude, shown in Fig.~\ref{fig:2}(c), present a rather homogeneous distribution, with shape in agreement with the electron microscopy characterization (Fig. \ref{fig:1}(f)), assuming that the left part of the crystal is not contributing to the diffraction pattern. Using the 3D phase map (Fig.~\ref{fig:2}(d)), 111 strain and lattice plane tilt were extracted \cite{Pateras15}, see Fig.~\ref{fig:2}(e) and \ref{fig:2}(f). They show that the lattice distortions originate mainly from a lattice tilt of about 0.2$^\circ$, of one part of the crystal with respect to the other part, while the strain fluctuations are rather weak. For the BCDI data set, several inversion attempts were performed without success (see Fig.~\ref{fig:2}(b), \ref{fig:2}(g), and \ref{fig:2}(h)), although some of these attempts included an educated support and/or an educated initial object guess, as described in Section S3 of the Supplemental Materials \cite{Supplemental_Materials}. This phase retrieval failure likely results from the presence of the strong lattice distortions and the associated phase gradient distribution.  Finally, the 3DBP reconstruction was used to estimate the 3D coherent diffraction pattern and compare it to the BCDI acquisition. The very good agreement observed between Fig.~\ref{fig:2}(i) and \ref{fig:2}(j) confirms that the 3DBP retrieved particle is also a good solution of the BCDI data set. 

\begin{figure}
    \centering
    \includegraphics[width=1.0\linewidth]{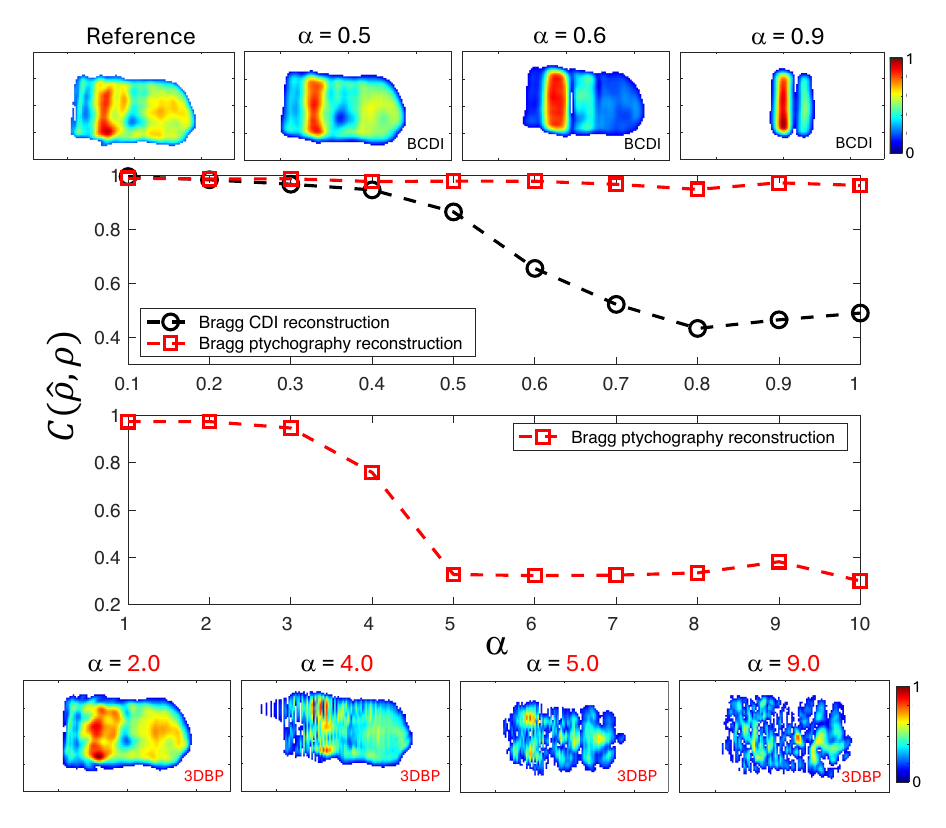}
    \caption{\textbf{Limits of BCDI and 3DBP methods assessed from numerical simulations}. Normalized cross-correlation values $C$ obtained by comparing the reconstruction result to the ground truth when the phase distortion parameter $\alpha$ increases from 0.1 to 1 (top plot) and from  1 to 10 (bottom plot). The $\alpha =1$ value corresponds to the distortion level of the highly distorted particle of Fig. \ref{fig:2}. Black circles and red squares correspond to BCDI and 3DBP results, respectively. The dashed lines are guides to the eye. On the top and bottom of the plot, 2D cross-sections of the BCDI and 3DBP retrieved amplitudes are shown, respectively, for some of the $\alpha$ values.}
    \label{fig:3}
\end{figure}

This experimental result serves as the starting point for investigating the performance limits of 3DBP and BCDI, which are subsequently explored using numerical approaches. The 3DBP reconstruction presented in Fig.~\ref{fig:2} is used as a reference object \textit{i.e.}, defining the reference amplitude $A_0(\mathbf{r})$ and  reference phase $\phi_0(\mathbf{r})$. From this reference, a series of numerical objects was generated to probe a broad range of lattice distortions. While the amplitude of these numerical objects was always set equal to $A_0(\mathbf{r})$, their phases were defined as $\alpha \phi_0(\mathbf{r})$, where $\alpha$ varies between 0.1 and 10 and serves as a tuning parameter for the lattice distortions. These numerical objects were used to generate BCDI and 3DBP data sets with sampling, extent and intensity dynamics matching those of the experimental data sets. The numerical data sets were subsequently inverted using the BCDI and 3DBP inversion strategies described in the Appendix C. The quality of the reconstructions was then assessed using the absolute value of the normalized cross-correlation, defined as
\begin{equation}
C(\hat{\rho}, \rho) = \left| \frac{\int \hat{\rho}^*(r)\,\rho(r)\,dr}{\lVert \hat{\rho} \rVert \, \lVert \rho \rVert}\right|
\end{equation}
\noindent and previously used in \cite{Calvo24}. Here, $\rho$ denotes the complex-valued function of the nominal object, while $\hat{\rho}$ corresponds to the reconstructed object. The superscript $^*$ indicates complex conjugation. Accordingly, $C = 1$ indicates perfect agreement between the retrieved and nominal objects, whereas smaller values of $C$ reflect increasing discrepancies and thus signal a failure of the inversion process.
Results obtained for $\alpha \leq 1$ are presented in Fig.~\ref{fig:3}, in black. BCDI successfully reconstructs the object for $\alpha \leq 0.5$; however, the agreement between the retrieved and nominal objects progressively degrades as $\alpha$ increases. This degradation is further illustrated by cross-sections of the retrieved amplitude shown at the top of Fig.~\ref{fig:3}, which reveal a marked loss of reconstruction quality for $\alpha \geq 0.6$. In contrast, the 3DBP approach yields fully satisfactory reconstructions over this range of $\alpha$ values. The behavior of the 3DBP method for larger distortion ranges ($1 \leq \alpha \leq 10$) is shown in Fig.~\ref{fig:3}, in red. These results indicate that the inversion process begins to fail for $\alpha > 3$, corresponding to a limit approximately six times higher than that observed for BCDI. For $\alpha = 3$, the phase gradient reaches approximately 0.12 rad/nm (or 2.1 rad/pixel), while the phase range is on the order of $7.2 \times 2\pi$ radians over 0.4 $\mu$m.

Another example of 3DBP reconstruction is presented in the Supplemental Materials, section S4 \cite{Supplemental_Materials}. While the particle was nano-indented prior to the 3DBP acquisition and exhibits a total phase range of approximately $11 \times 2\pi$, successful 3DBP reconstruction is still obtained (see also Supplemental Materials, section S5 \cite{Supplemental_Materials} for additional discussion on the retrieved image quality, amplitude and phase.)

\textit{Discussions and conclusions} - 
We experimentally and numerically show that 3DBP is suitable for imaging isolated crystalline particles. For highly distorted crystals, 3DBP tolerates lattice distortions more than six times larger than BCDI, while also providing improved reconstruction accuracy, as demonstrated for the weakly distorted crystal.

Nanoscale imaging of highly distorted finite-sized crystals is a long-standing challenge in microscopy. While BCDI offers 3D capability and \textit{in situ} compatibility, it is unable to invert the diffraction pattern of highly distorted crystals, \textit{i.e.}, crystals exhibiting strongly inhomogeneous phase distributions. 
Several approaches have been explored to mitigate this issue: Combining different Bragg reflection measurements \cite{Newton_multiple_2020}, using deep-learning strategies \cite{Masto_deep_learning_2026}, introducing phase modulators positioned before \cite{Calvo24} or after \cite{Zhao_modulator_2025} the sample. This often comes at the expense of experimental set-up and measurement strategy complexities.
In contrast, the present results show that 3DBP reliably reconstructs both weakly and highly distorted micro-crystals, making \textit{in situ} monitoring of developing lattice distortions feasible.

A clear understanding of the origin of the BCDI limitation is still lacking. Empirically, a correlation is observed between the failure of BCDI and the distortion of the diffraction pattern, the latter being defined as the discrepancy between the measured pattern and that expected solely from the particle morphology (or support). The existence and level of this threshold, and how it relates to the phase distribution within the particle and to the dynamical range of the data, remain open questions. A general consensus seems to recognize the difficulty to apply BCDI methods for object with phase range larger than $\pi$ \cite{Cha_2010, newton_prb_10, Huang_prb_2011} or $2\pi$ \cite{Pavlov_SR_2017, Zhao_modulator_2025} and the need to develop inversion methods to overcome this limit. In the presented work, 3DBP performs reliably until a phase range of $7.2 \times 2\pi$  and even $11 \times  2\pi$. These exceed the values of $4 \times 2\pi$ achieved recently \cite{Zhao_modulator_2025}. Moreover, 3DBP makes efficient use of the incoming photon flux by allowing the use of focused beams with focal spots comparable to or slightly larger than the particle size, in contrast to the plane-wave illumination required in BCDI (Supplemental Materials, section S2 \cite{Supplemental_Materials}). This aspect becomes particularly relevant for the investigation of particles smaller than 50 nm, such as catalytic particles \cite{Sun_review_2024}.

For \textit{in situ} monitoring of micro-crystal deformation under realistic operating conditions, several practical constraints must be considered. 
Besides limiting the acquisition time and using on-the-fly acquisitions, using an over-focused probe with a size comparable to that of the particle, as demonstrated here, minimizes the number of required translation steps. It also reduces the sensitivity of the experiment to vibrations and positioning errors, compared to acquisitions performed with a small spot size \cite{Li22}. 
Finally, the use of structured illumination is expected to help reducing the number of steps along the rocking curve direction \cite{Calvo24}. 
In parallel, combining 3DBP with \textit{operando} experiments may be challenging with respect to sample stability and integrity of its environment. These could be mitigated by translating the optics instead of the sample \cite{Hruszkewycz_ferro_2013} or introducing a phase modulator at the lens aperture and scanning it during the acquisition to ensure data diversity.

In summary, 3DBP extends coherent Bragg microscopy into the regime of strongly inhomogeneous strain, allowing quantitative reconstruction of the 3D displacement field in finite-sized crystals. This opens a direct experimental route to probe crystalline deformation under realistic operating conditions.

\textit{Acknowledgments} - The authors thank Dr. Darren Batey for his help with the experiments and Dr. Matt Spink for taking the SEM images at Diamond Light Source. We thank Nicholas Phillips and Annika F. Möslein for preparation of the nano-indented Au micro-crystal. DY and FH acknowledge funding from the European Research Council under the European Union’s Horizon 2020 research and innovation program (grant agreement No 714697). The authors also acknowledge Diamond Light Source for time on Beamline I13-1 under proposal numbers MG28441-2 and CM33893-2.

\textit{Data availability} - The data that support the findings of this article are openly available at [repository link].


\begin{thebibliography}{42}%
\makeatletter
\providecommand \@ifxundefined [1]{%
 \@ifx{#1\undefined}
}%
\providecommand \@ifnum [1]{%
 \ifnum #1\expandafter \@firstoftwo
 \else \expandafter \@secondoftwo
 \fi
}%
\providecommand \@ifx [1]{%
 \ifx #1\expandafter \@firstoftwo
 \else \expandafter \@secondoftwo
 \fi
}%
\providecommand \natexlab [1]{#1}%
\providecommand \enquote  [1]{``#1''}%
\providecommand \bibnamefont  [1]{#1}%
\providecommand \bibfnamefont [1]{#1}%
\providecommand \citenamefont [1]{#1}%
\providecommand \href@noop [0]{\@secondoftwo}%
\providecommand \href [0]{\begingroup \@sanitize@url \@href}%
\providecommand \@href[1]{\@@startlink{#1}\@@href}%
\providecommand \@@href[1]{\endgroup#1\@@endlink}%
\providecommand \@sanitize@url [0]{\catcode `\\12\catcode `\$12\catcode `\&12\catcode `\#12\catcode `\^12\catcode `\_12\catcode `\%12\relax}%
\providecommand \@@startlink[1]{}%
\providecommand \@@endlink[0]{}%
\providecommand \url  [0]{\begingroup\@sanitize@url \@url }%
\providecommand \@url [1]{\endgroup\@href {#1}{\urlprefix }}%
\providecommand \urlprefix  [0]{URL }%
\providecommand \Eprint [0]{\href }%
\providecommand \doibase [0]{https://doi.org/}%
\providecommand \selectlanguage [0]{\@gobble}%
\providecommand \bibinfo  [0]{\@secondoftwo}%
\providecommand \bibfield  [0]{\@secondoftwo}%
\providecommand \translation [1]{[#1]}%
\providecommand \BibitemOpen [0]{}%
\providecommand \bibitemStop [0]{}%
\providecommand \bibitemNoStop [0]{.\EOS\space}%
\providecommand \EOS [0]{\spacefactor3000\relax}%
\providecommand \BibitemShut  [1]{\csname bibitem#1\endcsname}%
\let\auto@bib@innerbib\@empty
\bibitem [{\citenamefont {Ostovar}\ \emph {et~al.}(2024)\citenamefont {Ostovar}, \citenamefont {Lee}, \citenamefont {Mehmood}, \citenamefont {Farrell}, \citenamefont {Searles}, \citenamefont {Bourgeois}, \citenamefont {Chiang}, \citenamefont {Misiura}, \citenamefont {Gross}, \citenamefont {Al-Zubeidi}, \citenamefont {Dionne}, \citenamefont {Landes}, \citenamefont {Zanni}, \citenamefont {Levine},\ and\ \citenamefont {Link}}]{Ostovar_sciadv_2024}%
  \BibitemOpen
  \bibfield  {author} {\bibinfo {author} {\bibfnamefont {B.}~\bibnamefont {Ostovar}}, \bibinfo {author} {\bibfnamefont {S.~A.}\ \bibnamefont {Lee}}, \bibinfo {author} {\bibfnamefont {A.}~\bibnamefont {Mehmood}}, \bibinfo {author} {\bibfnamefont {K.}~\bibnamefont {Farrell}}, \bibinfo {author} {\bibfnamefont {E.~K.}\ \bibnamefont {Searles}}, \bibinfo {author} {\bibfnamefont {B.}~\bibnamefont {Bourgeois}}, \bibinfo {author} {\bibfnamefont {W.-Y.}\ \bibnamefont {Chiang}}, \bibinfo {author} {\bibfnamefont {A.}~\bibnamefont {Misiura}}, \bibinfo {author} {\bibfnamefont {N.}~\bibnamefont {Gross}}, \bibinfo {author} {\bibfnamefont {A.}~\bibnamefont {Al-Zubeidi}}, \bibinfo {author} {\bibfnamefont {J.~A.}\ \bibnamefont {Dionne}}, \bibinfo {author} {\bibfnamefont {C.~F.}\ \bibnamefont {Landes}}, \bibinfo {author} {\bibfnamefont {M.}~\bibnamefont {Zanni}}, \bibinfo {author} {\bibfnamefont {B.~G.}\ \bibnamefont {Levine}},\ and\ \bibinfo {author} {\bibfnamefont {S.}~\bibnamefont {Link}},\ }\bibfield  {title} {\bibinfo
  {title} {The role of the plasmon in interfacial charge transfer},\ }\href {https://doi.org/10.1126/sciadv.adp3353} {\bibfield  {journal} {\bibinfo  {journal} {Science Advances}\ }\textbf {\bibinfo {volume} {10}},\ \bibinfo {pages} {eadp3353} (\bibinfo {year} {2024})}\BibitemShut {NoStop}%
\bibitem [{\citenamefont {Uebe}\ and\ \citenamefont {Schüler}(2016)}]{uebe_microbiol_2016}%
  \BibitemOpen
  \bibfield  {author} {\bibinfo {author} {\bibfnamefont {R.}~\bibnamefont {Uebe}}\ and\ \bibinfo {author} {\bibfnamefont {D.}~\bibnamefont {Schüler}},\ }\bibfield  {title} {\bibinfo {title} {Magnetosome biogenesis in magnetotactic bacteria},\ }\href {https://doi.org/10.1038/nrmicro.2016.99} {\bibfield  {journal} {\bibinfo  {journal} {Nature Reviews Microbiology}\ }\textbf {\bibinfo {volume} {14}},\ \bibinfo {pages} {621–637} (\bibinfo {year} {2016})}\BibitemShut {NoStop}%
\bibitem [{\citenamefont {Pei}\ \emph {et~al.}(2025)\citenamefont {Pei}, \citenamefont {Ringe}, \citenamefont {Chang}, \citenamefont {Harrison}, \citenamefont {Xue},\ and\ \citenamefont {Williams}}]{Pei_communications_2025}%
  \BibitemOpen
  \bibfield  {author} {\bibinfo {author} {\bibfnamefont {Z.}~\bibnamefont {Pei}}, \bibinfo {author} {\bibfnamefont {E.}~\bibnamefont {Ringe}}, \bibinfo {author} {\bibfnamefont {L.}~\bibnamefont {Chang}}, \bibinfo {author} {\bibfnamefont {R.~J.}\ \bibnamefont {Harrison}}, \bibinfo {author} {\bibfnamefont {P.}~\bibnamefont {Xue}},\ and\ \bibinfo {author} {\bibfnamefont {W.}~\bibnamefont {Williams}},\ }\bibfield  {title} {\bibinfo {title} {Three-dimensional structure, crystallography, and magnetism of giant magnetofossils},\ }\href {https://doi.org/10.1038/s43247-025-02396-} {\bibfield  {journal} {\bibinfo  {journal} {Communications Earth and Environment}\ }\textbf {\bibinfo {volume} {6}},\ \bibinfo {pages} {410} (\bibinfo {year} {2025})}\BibitemShut {NoStop}%
\bibitem [{\citenamefont {Yu}\ \emph {et~al.}(2015)\citenamefont {Yu}, \citenamefont {Kim}, \citenamefont {Shapiro}, \citenamefont {Farmand}, \citenamefont {Qian}, \citenamefont {Tyliszczak}, \citenamefont {Kilcoyne}, \citenamefont {Celestre}, \citenamefont {Marchesini}, \citenamefont {Joseph}, \citenamefont {Denes}, \citenamefont {Warwick}, \citenamefont {Strobridge}, \citenamefont {Grey}, \citenamefont {Padmore}, \citenamefont {Meng}, \citenamefont {Kostecki},\ and\ \citenamefont {Cabana}}]{Yu_nanolett_2015}%
  \BibitemOpen
  \bibfield  {author} {\bibinfo {author} {\bibfnamefont {Y.-S.}\ \bibnamefont {Yu}}, \bibinfo {author} {\bibfnamefont {C.}~\bibnamefont {Kim}}, \bibinfo {author} {\bibfnamefont {D.~A.}\ \bibnamefont {Shapiro}}, \bibinfo {author} {\bibfnamefont {M.}~\bibnamefont {Farmand}}, \bibinfo {author} {\bibfnamefont {D.}~\bibnamefont {Qian}}, \bibinfo {author} {\bibfnamefont {T.}~\bibnamefont {Tyliszczak}}, \bibinfo {author} {\bibfnamefont {A.~L.~D.}\ \bibnamefont {Kilcoyne}}, \bibinfo {author} {\bibfnamefont {R.}~\bibnamefont {Celestre}}, \bibinfo {author} {\bibfnamefont {S.}~\bibnamefont {Marchesini}}, \bibinfo {author} {\bibfnamefont {J.}~\bibnamefont {Joseph}}, \bibinfo {author} {\bibfnamefont {P.}~\bibnamefont {Denes}}, \bibinfo {author} {\bibfnamefont {T.}~\bibnamefont {Warwick}}, \bibinfo {author} {\bibfnamefont {F.~C.}\ \bibnamefont {Strobridge}}, \bibinfo {author} {\bibfnamefont {C.~P.}\ \bibnamefont {Grey}}, \bibinfo {author} {\bibfnamefont {H.}~\bibnamefont {Padmore}}, \bibinfo {author} {\bibfnamefont
  {Y.~S.}\ \bibnamefont {Meng}}, \bibinfo {author} {\bibfnamefont {R.}~\bibnamefont {Kostecki}},\ and\ \bibinfo {author} {\bibfnamefont {J.}~\bibnamefont {Cabana}},\ }\bibfield  {title} {\bibinfo {title} {Dependence on crystal size of the nanoscale chemical phase distribution and fracture in lixfepo4},\ }\href {https://doi.org/10.1021/acs.nanolett.5b01314} {\bibfield  {journal} {\bibinfo  {journal} {Nano Letters}\ }\textbf {\bibinfo {volume} {15}},\ \bibinfo {pages} {4282} (\bibinfo {year} {2015})}\BibitemShut {NoStop}%
\bibitem [{\citenamefont {Hughes}\ \emph {et~al.}(2022)\citenamefont {Hughes}, \citenamefont {Savitzky}, \citenamefont {Deng}, \citenamefont {Jin}, \citenamefont {Lomeli}, \citenamefont {Yu}, \citenamefont {Shapiro}, \citenamefont {Herring}, \citenamefont {Anapolsky}, \citenamefont {Chueh}, \citenamefont {Ophus},\ and\ \citenamefont {Minor}}]{Hughes_Materials_2022}%
  \BibitemOpen
  \bibfield  {author} {\bibinfo {author} {\bibfnamefont {L.}~\bibnamefont {Hughes}}, \bibinfo {author} {\bibfnamefont {B.~H.}\ \bibnamefont {Savitzky}}, \bibinfo {author} {\bibfnamefont {H.~D.}\ \bibnamefont {Deng}}, \bibinfo {author} {\bibfnamefont {N.~L.}\ \bibnamefont {Jin}}, \bibinfo {author} {\bibfnamefont {E.~G.}\ \bibnamefont {Lomeli}}, \bibinfo {author} {\bibfnamefont {Y.-S.}\ \bibnamefont {Yu}}, \bibinfo {author} {\bibfnamefont {D.~A.}\ \bibnamefont {Shapiro}}, \bibinfo {author} {\bibfnamefont {P.}~\bibnamefont {Herring}}, \bibinfo {author} {\bibfnamefont {A.}~\bibnamefont {Anapolsky}}, \bibinfo {author} {\bibfnamefont {W.~C.}\ \bibnamefont {Chueh}}, \bibinfo {author} {\bibfnamefont {C.}~\bibnamefont {Ophus}},\ and\ \bibinfo {author} {\bibfnamefont {A.~M.}\ \bibnamefont {Minor}},\ }\bibfield  {title} {\bibinfo {title} {Correlative analysis of structure and chemistry of lixfepo4 platelets using 4d-stem and x-ray ptychography},\ }\href {https://doi.org/https://doi.org/10.1016/j.mattod.2021.10.031}
  {\bibfield  {journal} {\bibinfo  {journal} {Materials Today}\ }\textbf {\bibinfo {volume} {52}},\ \bibinfo {pages} {102} (\bibinfo {year} {2022})}\BibitemShut {NoStop}%
\bibitem [{\citenamefont {Yang}\ \emph {et~al.}(2025)\citenamefont {Yang}, \citenamefont {Seif}, \citenamefont {He}, \citenamefont {Song}, \citenamefont {Morez}, \citenamefont {de~Jager}, \citenamefont {Nykypanchuk}, \citenamefont {Harder}, \citenamefont {Cha}, \citenamefont {Tarleton}, \citenamefont {Robinson},\ and\ \citenamefont {Hofmann}}]{Yang_advmater_2025}%
  \BibitemOpen
  \bibfield  {author} {\bibinfo {author} {\bibfnamefont {D.}~\bibnamefont {Yang}}, \bibinfo {author} {\bibfnamefont {M.}~\bibnamefont {Seif}}, \bibinfo {author} {\bibfnamefont {G.}~\bibnamefont {He}}, \bibinfo {author} {\bibfnamefont {K.}~\bibnamefont {Song}}, \bibinfo {author} {\bibfnamefont {A.}~\bibnamefont {Morez}}, \bibinfo {author} {\bibfnamefont {B.}~\bibnamefont {de~Jager}}, \bibinfo {author} {\bibfnamefont {D.}~\bibnamefont {Nykypanchuk}}, \bibinfo {author} {\bibfnamefont {R.~J.}\ \bibnamefont {Harder}}, \bibinfo {author} {\bibfnamefont {W.}~\bibnamefont {Cha}}, \bibinfo {author} {\bibfnamefont {E.}~\bibnamefont {Tarleton}}, \bibinfo {author} {\bibfnamefont {I.~K.}\ \bibnamefont {Robinson}},\ and\ \bibinfo {author} {\bibfnamefont {F.}~\bibnamefont {Hofmann}},\ }\bibfield  {title} {\bibinfo {title} {Direct imaging of hydrogen-driven dislocation and strain field evolution in a stainless steel grain},\ }\href {https://doi.org/https://doi.org/10.1002/adma.202500221} {\bibfield  {journal} {\bibinfo
  {journal} {Advanced Materials}\ }\textbf {\bibinfo {volume} {37}},\ \bibinfo {pages} {e00221} (\bibinfo {year} {2025})}\BibitemShut {NoStop}%
\bibitem [{\citenamefont {Perrut}\ \emph {et~al.}(2018)\citenamefont {Perrut}, \citenamefont {Caron}, \citenamefont {Thomas},\ and\ \citenamefont {Couret}}]{Perrut_CRPHys_2018}%
  \BibitemOpen
  \bibfield  {author} {\bibinfo {author} {\bibfnamefont {M.}~\bibnamefont {Perrut}}, \bibinfo {author} {\bibfnamefont {P.}~\bibnamefont {Caron}}, \bibinfo {author} {\bibfnamefont {M.}~\bibnamefont {Thomas}},\ and\ \bibinfo {author} {\bibfnamefont {A.}~\bibnamefont {Couret}},\ }\bibfield  {title} {\bibinfo {title} {High temperature materials for aerospace applications: Ni-based superalloys and gamma-tial alloys},\ }\href {https://doi.org/https://doi.org/10.1016/j.crhy.2018.10.002} {\bibfield  {journal} {\bibinfo  {journal} {Comptes Rendus Physique}\ }\textbf {\bibinfo {volume} {19}},\ \bibinfo {pages} {657} (\bibinfo {year} {2018})}\BibitemShut {NoStop}%
\bibitem [{\citenamefont {Jo}\ \emph {et~al.}(2022)\citenamefont {Jo}, \citenamefont {Wi}, \citenamefont {Lee}, \citenamefont {Kwon}, \citenamefont {Jeong}, \citenamefont {Lee}, \citenamefont {Baik}, \citenamefont {Pattison}, \citenamefont {Theis}, \citenamefont {Ophus}, \citenamefont {Lee}, \citenamefont {Ryu},\ and\ \citenamefont {Yang}}]{Jo_NatCommun_2022}%
  \BibitemOpen
  \bibfield  {author} {\bibinfo {author} {\bibfnamefont {H.}~\bibnamefont {Jo}}, \bibinfo {author} {\bibfnamefont {D.~H.}\ \bibnamefont {Wi}}, \bibinfo {author} {\bibfnamefont {T.}~\bibnamefont {Lee}}, \bibinfo {author} {\bibfnamefont {Y.}~\bibnamefont {Kwon}}, \bibinfo {author} {\bibfnamefont {C.}~\bibnamefont {Jeong}}, \bibinfo {author} {\bibfnamefont {J.}~\bibnamefont {Lee}}, \bibinfo {author} {\bibfnamefont {H.}~\bibnamefont {Baik}}, \bibinfo {author} {\bibfnamefont {A.~J.}\ \bibnamefont {Pattison}}, \bibinfo {author} {\bibfnamefont {W.}~\bibnamefont {Theis}}, \bibinfo {author} {\bibfnamefont {P.}~\bibnamefont {Ophus}, \bibfnamefont {Colin~Ercius}}, \bibinfo {author} {\bibfnamefont {Y.-L.}\ \bibnamefont {Lee}}, \bibinfo {author} {\bibfnamefont {S.~W.}\ \bibnamefont {Ryu}, \bibfnamefont {Seunghwa~Han}},\ and\ \bibinfo {author} {\bibfnamefont {Y.}~\bibnamefont {Yang}},\ }\bibfield  {title} {\bibinfo {title} {Direct strain correlations at the single-atom level in three-dimensional core-shell interface
  structures},\ }\href {https://doi.org/10.1038/s41467-022-33236-6} {\bibfield  {journal} {\bibinfo  {journal} {Nature Communications}\ }\textbf {\bibinfo {volume} {13}},\ \bibinfo {pages} {5957} (\bibinfo {year} {2022})}\BibitemShut {NoStop}%
\bibitem [{\citenamefont {Perx{\'e}s~Perich}\ \emph {et~al.}(2025)\citenamefont {Perx{\'e}s~Perich}, \citenamefont {Lankman}, \citenamefont {Keijzer},\ and\ \citenamefont {van~der Hoeven}}]{Perich_NanoLett_2025}%
  \BibitemOpen
  \bibfield  {author} {\bibinfo {author} {\bibfnamefont {M.}~\bibnamefont {Perx{\'e}s~Perich}}, \bibinfo {author} {\bibfnamefont {J.-W.}\ \bibnamefont {Lankman}}, \bibinfo {author} {\bibfnamefont {C.~J.}\ \bibnamefont {Keijzer}},\ and\ \bibinfo {author} {\bibfnamefont {J.~E.~S.}\ \bibnamefont {van~der Hoeven}},\ }\bibfield  {title} {\bibinfo {title} {In situ gas-phase 4d-stem for strain mapping during hydride formation in palladium nanocubes},\ }\href {https://doi.org/10.1021/acs.nanolett.5c00702} {\bibfield  {journal} {\bibinfo  {journal} {Nano Letters}\ }\textbf {\bibinfo {volume} {25}},\ \bibinfo {pages} {5444} (\bibinfo {year} {2025})}\BibitemShut {NoStop}%
\bibitem [{\citenamefont {Shukla}\ \emph {et~al.}(2025)\citenamefont {Shukla}, \citenamefont {Yildirim}, \citenamefont {Ball}, \citenamefont {Detlefs}, \citenamefont {Cretton}, \citenamefont {Sarkis}, \citenamefont {Bella}, \citenamefont {Ludwig}, \citenamefont {Zhang}, ,\ and\ \citenamefont {Henningsson}}]{shukla_2025}%
  \BibitemOpen
  \bibfield  {author} {\bibinfo {author} {\bibfnamefont {A.}~\bibnamefont {Shukla}}, \bibinfo {author} {\bibfnamefont {C.}~\bibnamefont {Yildirim}}, \bibinfo {author} {\bibfnamefont {J.~A.~D.}\ \bibnamefont {Ball}}, \bibinfo {author} {\bibfnamefont {C.}~\bibnamefont {Detlefs}}, \bibinfo {author} {\bibfnamefont {A.~A.~W.}\ \bibnamefont {Cretton}}, \bibinfo {author} {\bibfnamefont {M.}~\bibnamefont {Sarkis}}, \bibinfo {author} {\bibfnamefont {M.~L.}\ \bibnamefont {Bella}}, \bibinfo {author} {\bibfnamefont {W.}~\bibnamefont {Ludwig}}, \bibinfo {author} {\bibfnamefont {Y.}~\bibnamefont {Zhang}}, ,\ and\ \bibinfo {author} {\bibfnamefont {N.~A.}\ \bibnamefont {Henningsson}},\ }\href {https://arxiv.org/abs/2508.17897} {\bibinfo {title} {Bridging grain mapping and dark field x-ray microscopy for multiscale diffraction imaging}} (\bibinfo {year} {2025})\BibitemShut {NoStop}%
\bibitem [{\citenamefont {Pfeifer}\ \emph {et~al.}(2006)\citenamefont {Pfeifer}, \citenamefont {Williams}, \citenamefont {Vartanyants}, \citenamefont {Harder},\ and\ \citenamefont {Robinson}}]{pfeifer06}%
  \BibitemOpen
  \bibfield  {author} {\bibinfo {author} {\bibfnamefont {M.~A.}\ \bibnamefont {Pfeifer}}, \bibinfo {author} {\bibfnamefont {G.~J.}\ \bibnamefont {Williams}}, \bibinfo {author} {\bibfnamefont {I.~A.}\ \bibnamefont {Vartanyants}}, \bibinfo {author} {\bibfnamefont {R.}~\bibnamefont {Harder}},\ and\ \bibinfo {author} {\bibfnamefont {I.~K.}\ \bibnamefont {Robinson}},\ }\bibfield  {title} {\bibinfo {title} {Three-dimensional mapping of a deformation field inside a nanocrystal},\ }\href@noop {} {\bibfield  {journal} {\bibinfo  {journal} {Nature}\ }\textbf {\bibinfo {volume} {442}},\ \bibinfo {pages} {63} (\bibinfo {year} {2006})}\BibitemShut {NoStop}%
\bibitem [{\citenamefont {Hofmann}\ \emph {et~al.}(2017)\citenamefont {Hofmann}, \citenamefont {Phillips}, \citenamefont {Harder}, \citenamefont {Liu}, \citenamefont {Clark}, \citenamefont {Robinson},\ and\ \citenamefont {Abbey}}]{Hofmann_multi_Bragg_2017}%
  \BibitemOpen
  \bibfield  {author} {\bibinfo {author} {\bibfnamefont {F.}~\bibnamefont {Hofmann}}, \bibinfo {author} {\bibfnamefont {N.~W.}\ \bibnamefont {Phillips}}, \bibinfo {author} {\bibfnamefont {R.~J.}\ \bibnamefont {Harder}}, \bibinfo {author} {\bibfnamefont {W.}~\bibnamefont {Liu}}, \bibinfo {author} {\bibfnamefont {J.~N.}\ \bibnamefont {Clark}}, \bibinfo {author} {\bibfnamefont {I.~K.}\ \bibnamefont {Robinson}},\ and\ \bibinfo {author} {\bibfnamefont {B.}~\bibnamefont {Abbey}},\ }\bibfield  {title} {\bibinfo {title} {Micro-beam laue alignment of multi-reflection bragg coherent diffraction imaging measurements},\ }\href {https://doi.org/10.1107/S1600577517009183} {\bibfield  {journal} {\bibinfo  {journal} {J Synchrotron Radiat}\ }\textbf {\bibinfo {volume} {24}},\ \bibinfo {pages} {1048} (\bibinfo {year} {2017})}\BibitemShut {NoStop}%
\bibitem [{\citenamefont {Cha}\ \emph {et~al.}(2016)\citenamefont {Cha}, \citenamefont {Ulvestad}, \citenamefont {Allain}, \citenamefont {Chamard}, \citenamefont {Harder}, \citenamefont {Leake}, \citenamefont {Maser}, \citenamefont {Fuoss},\ and\ \citenamefont {Hruszkewycz}}]{Cha_energy_2015}%
  \BibitemOpen
  \bibfield  {author} {\bibinfo {author} {\bibfnamefont {W.}~\bibnamefont {Cha}}, \bibinfo {author} {\bibfnamefont {A.}~\bibnamefont {Ulvestad}}, \bibinfo {author} {\bibfnamefont {M.}~\bibnamefont {Allain}}, \bibinfo {author} {\bibfnamefont {V.}~\bibnamefont {Chamard}}, \bibinfo {author} {\bibfnamefont {R.}~\bibnamefont {Harder}}, \bibinfo {author} {\bibfnamefont {S.~J.}\ \bibnamefont {Leake}}, \bibinfo {author} {\bibfnamefont {J.}~\bibnamefont {Maser}}, \bibinfo {author} {\bibfnamefont {P.~H.}\ \bibnamefont {Fuoss}},\ and\ \bibinfo {author} {\bibfnamefont {S.~O.}\ \bibnamefont {Hruszkewycz}},\ }\bibfield  {title} {\bibinfo {title} {Three dimensional variable-wavelength x-ray bragg coherent diffraction imaging},\ }\href {https://doi.org/10.1103/PhysRevLett.117.225501} {\bibfield  {journal} {\bibinfo  {journal} {Phys. Rev. Lett.}\ }\textbf {\bibinfo {volume} {117}},\ \bibinfo {pages} {225501} (\bibinfo {year} {2016})}\BibitemShut {NoStop}%
\bibitem [{\citenamefont {Grimes}\ \emph {et~al.}(2024)\citenamefont {Grimes}, \citenamefont {Atlan}, \citenamefont {Chatelier}, \citenamefont {Bellec}, \citenamefont {Olson}, \citenamefont {Simonne}, \citenamefont {Levi}, \citenamefont {Sch{\"u}lli}, \citenamefont {Leake}, \citenamefont {Rabkin}, \citenamefont {Eymery},\ and\ \citenamefont {Richard}}]{Simonne_catalysis_2024}%
  \BibitemOpen
  \bibfield  {author} {\bibinfo {author} {\bibfnamefont {M.}~\bibnamefont {Grimes}}, \bibinfo {author} {\bibfnamefont {C.}~\bibnamefont {Atlan}}, \bibinfo {author} {\bibfnamefont {C.}~\bibnamefont {Chatelier}}, \bibinfo {author} {\bibfnamefont {E.}~\bibnamefont {Bellec}}, \bibinfo {author} {\bibfnamefont {K.}~\bibnamefont {Olson}}, \bibinfo {author} {\bibfnamefont {D.}~\bibnamefont {Simonne}}, \bibinfo {author} {\bibfnamefont {M.}~\bibnamefont {Levi}}, \bibinfo {author} {\bibfnamefont {T.~U.}\ \bibnamefont {Sch{\"u}lli}}, \bibinfo {author} {\bibfnamefont {S.~J.}\ \bibnamefont {Leake}}, \bibinfo {author} {\bibfnamefont {E.}~\bibnamefont {Rabkin}}, \bibinfo {author} {\bibfnamefont {J.}~\bibnamefont {Eymery}},\ and\ \bibinfo {author} {\bibfnamefont {M.-I.}\ \bibnamefont {Richard}},\ }\bibfield  {title} {\bibinfo {title} {Capturing catalyst strain dynamics during operando co oxidation},\ }\href {https://doi.org/10.1021/acsnano.4c04127} {\bibfield  {journal} {\bibinfo  {journal} {ACS Nano}\ }\textbf {\bibinfo
  {volume} {18}},\ \bibinfo {pages} {19608} (\bibinfo {year} {2024})}\BibitemShut {NoStop}%
\bibitem [{\citenamefont {Atlan}\ \emph {et~al.}(2023)\citenamefont {Atlan}, \citenamefont {Chatelier}, \citenamefont {Martens}, \citenamefont {Dupraz}, \citenamefont {Viola}, \citenamefont {Li}, \citenamefont {Gao}, \citenamefont {Leake}, \citenamefont {Schülli}, \citenamefont {Eymery}, \citenamefont {Maillard},\ and\ \citenamefont {Richard}}]{Atlan_electrochem_BCDI_2023}%
  \BibitemOpen
  \bibfield  {author} {\bibinfo {author} {\bibfnamefont {C.}~\bibnamefont {Atlan}}, \bibinfo {author} {\bibfnamefont {C.}~\bibnamefont {Chatelier}}, \bibinfo {author} {\bibfnamefont {I.}~\bibnamefont {Martens}}, \bibinfo {author} {\bibfnamefont {M.}~\bibnamefont {Dupraz}}, \bibinfo {author} {\bibfnamefont {A.}~\bibnamefont {Viola}}, \bibinfo {author} {\bibfnamefont {N.}~\bibnamefont {Li}}, \bibinfo {author} {\bibfnamefont {L.}~\bibnamefont {Gao}}, \bibinfo {author} {\bibfnamefont {S.~J.}\ \bibnamefont {Leake}}, \bibinfo {author} {\bibfnamefont {T.~U.}\ \bibnamefont {Schülli}}, \bibinfo {author} {\bibfnamefont {J.}~\bibnamefont {Eymery}}, \bibinfo {author} {\bibfnamefont {F.}~\bibnamefont {Maillard}},\ and\ \bibinfo {author} {\bibfnamefont {M.-I.}\ \bibnamefont {Richard}},\ }\bibfield  {title} {\bibinfo {title} {Imaging the strain evolution of a platinum nanoparticle under electrochemical control},\ }\href {https://doi.org/10.1038/s41563-023-01528-x} {\bibfield  {journal} {\bibinfo  {journal} {Nat. Mater.}\
  }\textbf {\bibinfo {volume} {22}},\ \bibinfo {pages} {754–761} (\bibinfo {year} {2023})}\BibitemShut {NoStop}%
\bibitem [{\citenamefont {Yehya}\ \emph {et~al.}(2024)\citenamefont {Yehya}, \citenamefont {Cornelius}, \citenamefont {Richard}, \citenamefont {Berenguer}, \citenamefont {Levi}, \citenamefont {Rabkin}, \citenamefont {Thomas},\ and\ \citenamefont {Labat}}]{Yehya_nanoindentBCDI_2024}%
  \BibitemOpen
  \bibfield  {author} {\bibinfo {author} {\bibfnamefont {S.}~\bibnamefont {Yehya}}, \bibinfo {author} {\bibfnamefont {T.~W.}\ \bibnamefont {Cornelius}}, \bibinfo {author} {\bibfnamefont {M.-I.}\ \bibnamefont {Richard}}, \bibinfo {author} {\bibfnamefont {F.}~\bibnamefont {Berenguer}}, \bibinfo {author} {\bibfnamefont {M.}~\bibnamefont {Levi}}, \bibinfo {author} {\bibfnamefont {E.}~\bibnamefont {Rabkin}}, \bibinfo {author} {\bibfnamefont {O.}~\bibnamefont {Thomas}},\ and\ \bibinfo {author} {\bibfnamefont {S.}~\bibnamefont {Labat}},\ }\bibfield  {title} {\bibinfo {title} {In situ three-dimensional observation of plasticity onset in a pt nanoparticle},\ }\href {https://doi.org/10.1039/D4NR02634A} {\bibfield  {journal} {\bibinfo  {journal} {Nanoscale}\ }\textbf {\bibinfo {volume} {16}},\ \bibinfo {pages} {20670} (\bibinfo {year} {2024})}\BibitemShut {NoStop}%
\bibitem [{\citenamefont {Clark}\ \emph {et~al.}(2015)\citenamefont {Clark}, \citenamefont {Ihli}, \citenamefont {Schenk}, \citenamefont {Kim}, \citenamefont {Kulak}, \citenamefont {Campbell}, \citenamefont {Nisbet}, \citenamefont {Meldrum},\ and\ \citenamefont {Robinson}}]{Clark_dissolution_2015}%
  \BibitemOpen
  \bibfield  {author} {\bibinfo {author} {\bibfnamefont {J.~N.}\ \bibnamefont {Clark}}, \bibinfo {author} {\bibfnamefont {J.}~\bibnamefont {Ihli}}, \bibinfo {author} {\bibfnamefont {A.~S.}\ \bibnamefont {Schenk}}, \bibinfo {author} {\bibfnamefont {Y.-Y.}\ \bibnamefont {Kim}}, \bibinfo {author} {\bibfnamefont {A.~N.}\ \bibnamefont {Kulak}}, \bibinfo {author} {\bibfnamefont {J.~M.}\ \bibnamefont {Campbell}}, \bibinfo {author} {\bibfnamefont {G.}~\bibnamefont {Nisbet}}, \bibinfo {author} {\bibfnamefont {F.~C.}\ \bibnamefont {Meldrum}},\ and\ \bibinfo {author} {\bibfnamefont {I.~K.}\ \bibnamefont {Robinson}},\ }\bibfield  {title} {\bibinfo {title} {Three-dimensional imaging of dislocation propagation during crystal growth and dissolution},\ }\href {https://doi.org/10.1038/nmat4320} {\bibfield  {journal} {\bibinfo  {journal} {Nature Materials}\ }\textbf {\bibinfo {volume} {14}},\ \bibinfo {pages} {780} (\bibinfo {year} {2015})}\BibitemShut {NoStop}%
\bibitem [{\citenamefont {Cha}\ \emph {et~al.}(2010)\citenamefont {Cha}, \citenamefont {Song}, \citenamefont {Jeong}, \citenamefont {Harder}, \citenamefont {Yoon}, \citenamefont {Robinson},\ and\ \citenamefont {Kim}}]{Cha_2010}%
  \BibitemOpen
  \bibfield  {author} {\bibinfo {author} {\bibfnamefont {W.}~\bibnamefont {Cha}}, \bibinfo {author} {\bibfnamefont {S.}~\bibnamefont {Song}}, \bibinfo {author} {\bibfnamefont {N.~C.}\ \bibnamefont {Jeong}}, \bibinfo {author} {\bibfnamefont {R.}~\bibnamefont {Harder}}, \bibinfo {author} {\bibfnamefont {K.~B.}\ \bibnamefont {Yoon}}, \bibinfo {author} {\bibfnamefont {I.~K.}\ \bibnamefont {Robinson}},\ and\ \bibinfo {author} {\bibfnamefont {H.}~\bibnamefont {Kim}},\ }\bibfield  {title} {\bibinfo {title} {Exploration of crystal strains using coherent x-ray diffraction},\ }\href {https://doi.org/10.1088/1367-2630/12/3/035022} {\bibfield  {journal} {\bibinfo  {journal} {New Journal of Physics}\ }\textbf {\bibinfo {volume} {12}},\ \bibinfo {pages} {035022} (\bibinfo {year} {2010})}\BibitemShut {NoStop}%
\bibitem [{\citenamefont {Newton}\ \emph {et~al.}(2010)\citenamefont {Newton}, \citenamefont {Harder}, \citenamefont {Huang}, \citenamefont {Xiong},\ and\ \citenamefont {Robinson}}]{newton_prb_10}%
  \BibitemOpen
  \bibfield  {author} {\bibinfo {author} {\bibfnamefont {M.}~\bibnamefont {Newton}}, \bibinfo {author} {\bibfnamefont {R.}~\bibnamefont {Harder}}, \bibinfo {author} {\bibfnamefont {X.}~\bibnamefont {Huang}}, \bibinfo {author} {\bibfnamefont {G.}~\bibnamefont {Xiong}},\ and\ \bibinfo {author} {\bibfnamefont {I.}~\bibnamefont {Robinson}},\ }\bibfield  {title} {\bibinfo {title} {Phase retrieval of diffraction from highly strained crystals},\ }\href@noop {} {\bibfield  {journal} {\bibinfo  {journal} {Physical Review B}\ }\textbf {\bibinfo {volume} {82}},\ \bibinfo {pages} {165436} (\bibinfo {year} {2010})}\BibitemShut {NoStop}%
\bibitem [{\citenamefont {Huang}\ \emph {et~al.}(2011)\citenamefont {Huang}, \citenamefont {Harder}, \citenamefont {Xiong}, \citenamefont {Shi},\ and\ \citenamefont {Robinson}}]{Huang_prb_2011}%
  \BibitemOpen
  \bibfield  {author} {\bibinfo {author} {\bibfnamefont {X.}~\bibnamefont {Huang}}, \bibinfo {author} {\bibfnamefont {R.}~\bibnamefont {Harder}}, \bibinfo {author} {\bibfnamefont {G.}~\bibnamefont {Xiong}}, \bibinfo {author} {\bibfnamefont {X.}~\bibnamefont {Shi}},\ and\ \bibinfo {author} {\bibfnamefont {I.}~\bibnamefont {Robinson}},\ }\bibfield  {title} {\bibinfo {title} {Propagation uniqueness in three-dimensional coherent diffractive imaging},\ }\href {https://doi.org/10.1103/PhysRevB.83.224109} {\bibfield  {journal} {\bibinfo  {journal} {Phys. Rev. B}\ }\textbf {\bibinfo {volume} {83}},\ \bibinfo {pages} {224109} (\bibinfo {year} {2011})}\BibitemShut {NoStop}%
\bibitem [{\citenamefont {Pavlov}\ \emph {et~al.}(2017)\citenamefont {Pavlov}, \citenamefont {Punegov}, \citenamefont {Morgan}, \citenamefont {Schmalz},\ and\ \citenamefont {Paganin}}]{Pavlov_SR_2017}%
  \BibitemOpen
  \bibfield  {author} {\bibinfo {author} {\bibfnamefont {K.~M.}\ \bibnamefont {Pavlov}}, \bibinfo {author} {\bibfnamefont {V.~I.}\ \bibnamefont {Punegov}}, \bibinfo {author} {\bibfnamefont {K.~S.}\ \bibnamefont {Morgan}}, \bibinfo {author} {\bibfnamefont {G.}~\bibnamefont {Schmalz}},\ and\ \bibinfo {author} {\bibfnamefont {D.~M.}\ \bibnamefont {Paganin}},\ }\bibfield  {title} {\bibinfo {title} {Deterministic bragg coherent diffraction imaging},\ }\href {https://doi.org/10.1038/s41598-017-01164-x} {\bibfield  {journal} {\bibinfo  {journal} {Sci Rep}\ }\textbf {\bibinfo {volume} {7}},\ \bibinfo {pages} {1132} (\bibinfo {year} {2017})}\BibitemShut {NoStop}%
\bibitem [{\citenamefont {Zhao}\ \emph {et~al.}(2025)\citenamefont {Zhao}, \citenamefont {Bellec}, \citenamefont {Richard}, \citenamefont {Pithan}, \citenamefont {Vartanyants}, \citenamefont {Zhang}, \citenamefont {Sch\"ulli},\ and\ \citenamefont {Leake}}]{Zhao_modulator_2025}%
  \BibitemOpen
  \bibfield  {author} {\bibinfo {author} {\bibfnamefont {J.}~\bibnamefont {Zhao}}, \bibinfo {author} {\bibfnamefont {E.}~\bibnamefont {Bellec}}, \bibinfo {author} {\bibfnamefont {M.-I.}\ \bibnamefont {Richard}}, \bibinfo {author} {\bibfnamefont {L.}~\bibnamefont {Pithan}}, \bibinfo {author} {\bibfnamefont {I.~A.}\ \bibnamefont {Vartanyants}}, \bibinfo {author} {\bibfnamefont {F.}~\bibnamefont {Zhang}}, \bibinfo {author} {\bibfnamefont {T.}~\bibnamefont {Sch\"ulli}},\ and\ \bibinfo {author} {\bibfnamefont {S.~J.}\ \bibnamefont {Leake}},\ }\bibfield  {title} {\bibinfo {title} {Bragg coherent modulation imaging of highly strained nanocrystals},\ }\href {https://doi.org/10.1103/8vrs-hjqm} {\bibfield  {journal} {\bibinfo  {journal} {Phys. Rev. Lett.}\ }\textbf {\bibinfo {volume} {135}},\ \bibinfo {pages} {256101} (\bibinfo {year} {2025})}\BibitemShut {NoStop}%
\bibitem [{\citenamefont {Suzana}\ \emph {et~al.}(2024)\citenamefont {Suzana}, \citenamefont {Lee}, \citenamefont {Calvo-Almazán}, \citenamefont {Cha}, \citenamefont {Harder},\ and\ \citenamefont {Fenter}}]{Suzana_calcite_2024}%
  \BibitemOpen
  \bibfield  {author} {\bibinfo {author} {\bibfnamefont {A.~F.}\ \bibnamefont {Suzana}}, \bibinfo {author} {\bibfnamefont {S.~S.}\ \bibnamefont {Lee}}, \bibinfo {author} {\bibfnamefont {I.}~\bibnamefont {Calvo-Almazán}}, \bibinfo {author} {\bibfnamefont {W.}~\bibnamefont {Cha}}, \bibinfo {author} {\bibfnamefont {R.}~\bibnamefont {Harder}},\ and\ \bibinfo {author} {\bibfnamefont {P.}~\bibnamefont {Fenter}},\ }\bibfield  {title} {\bibinfo {title} {Visualizing the internal nanocrystallinity of calcite due to nonclassical crystallization by 3d coherent x-ray diffraction imaging},\ }\href {https://doi.org/10.1002/adma.202310672} {\bibfield  {journal} {\bibinfo  {journal} {Adv Mater.}\ }\textbf {\bibinfo {volume} {36}},\ \bibinfo {pages} {e2310672} (\bibinfo {year} {2024})}\BibitemShut {NoStop}%
\bibitem [{\citenamefont {Mastropietro}\ \emph {et~al.}(2017)\citenamefont {Mastropietro}, \citenamefont {Allain}, \citenamefont {Burghammer}, \citenamefont {Chevallard}, \citenamefont {Daillant}, \citenamefont {Duboisset}, \citenamefont {Godard}, \citenamefont {Guenoun}, \citenamefont {Nouet},\ and\ \citenamefont {Chamard}}]{mastropietro_NatMat_17}%
  \BibitemOpen
  \bibfield  {author} {\bibinfo {author} {\bibfnamefont {F.}~\bibnamefont {Mastropietro}}, \bibinfo {author} {\bibfnamefont {M.}~\bibnamefont {Allain}}, \bibinfo {author} {\bibfnamefont {M.}~\bibnamefont {Burghammer}}, \bibinfo {author} {\bibfnamefont {C.}~\bibnamefont {Chevallard}}, \bibinfo {author} {\bibfnamefont {J.}~\bibnamefont {Daillant}}, \bibinfo {author} {\bibfnamefont {J.}~\bibnamefont {Duboisset}}, \bibinfo {author} {\bibfnamefont {P.}~\bibnamefont {Godard}}, \bibinfo {author} {\bibfnamefont {P.}~\bibnamefont {Guenoun}}, \bibinfo {author} {\bibfnamefont {J.}~\bibnamefont {Nouet}},\ and\ \bibinfo {author} {\bibfnamefont {V.}~\bibnamefont {Chamard}},\ }\bibfield  {title} {\bibinfo {title} {Revealing crystalline domains into a mollusk \emph{single-crystalline} prism},\ }\href@noop {} {\bibfield  {journal} {\bibinfo  {journal} {Nature Materials}\ }\textbf {\bibinfo {volume} {16}},\ \bibinfo {pages} {946} (\bibinfo {year} {2017})}\BibitemShut {NoStop}%
\bibitem [{\citenamefont {Hill}\ \emph {et~al.}(2018)\citenamefont {Hill}, \citenamefont {Calvo-Almazan}, \citenamefont {Allain}, \citenamefont {Holt}, \citenamefont {Ulvestad}, \citenamefont {Treu}, \citenamefont {Koblmüller}, \citenamefont {Huang}, \citenamefont {Huang}, \citenamefont {Yan}, \citenamefont {Nazaretski}, \citenamefont {Chu}, \citenamefont {Stephenson}, \citenamefont {Chamard}, \citenamefont {Lauhon},\ and\ \citenamefont {Hruszkewycz}}]{Hill2018}%
  \BibitemOpen
  \bibfield  {author} {\bibinfo {author} {\bibfnamefont {M.~O.}\ \bibnamefont {Hill}}, \bibinfo {author} {\bibfnamefont {I.}~\bibnamefont {Calvo-Almazan}}, \bibinfo {author} {\bibfnamefont {M.}~\bibnamefont {Allain}}, \bibinfo {author} {\bibfnamefont {M.~V.}\ \bibnamefont {Holt}}, \bibinfo {author} {\bibfnamefont {A.}~\bibnamefont {Ulvestad}}, \bibinfo {author} {\bibfnamefont {J.}~\bibnamefont {Treu}}, \bibinfo {author} {\bibfnamefont {G.}~\bibnamefont {Koblmüller}}, \bibinfo {author} {\bibfnamefont {C.}~\bibnamefont {Huang}}, \bibinfo {author} {\bibfnamefont {X.}~\bibnamefont {Huang}}, \bibinfo {author} {\bibfnamefont {H.}~\bibnamefont {Yan}}, \bibinfo {author} {\bibfnamefont {E.}~\bibnamefont {Nazaretski}}, \bibinfo {author} {\bibfnamefont {Y.~S.}\ \bibnamefont {Chu}}, \bibinfo {author} {\bibfnamefont {G.~B.}\ \bibnamefont {Stephenson}}, \bibinfo {author} {\bibfnamefont {V.}~\bibnamefont {Chamard}}, \bibinfo {author} {\bibfnamefont {L.~J.}\ \bibnamefont {Lauhon}},\ and\ \bibinfo {author} {\bibfnamefont
  {S.~O.}\ \bibnamefont {Hruszkewycz}},\ }\bibfield  {title} {\bibinfo {title} {{Measuring Three-Dimensional Strain and Structural Defects in a Single InGaAs Nanowire Using Coherent X-ray Multiangle Bragg Projection Ptychography}},\ }\href {https://doi.org/10.1021/acs.nanolett.7b04024} {\bibfield  {journal} {\bibinfo  {journal} {Nano Lett.}\ }\textbf {\bibinfo {volume} {18}},\ \bibinfo {pages} {811} (\bibinfo {year} {2018})}\BibitemShut {NoStop}%
\bibitem [{\citenamefont {Li}\ \emph {et~al.}(2021)\citenamefont {Li}, \citenamefont {Phillips}, \citenamefont {Leake}, \citenamefont {Allain}, \citenamefont {Hofmann},\ and\ \citenamefont {Chamard}}]{li21}%
  \BibitemOpen
  \bibfield  {author} {\bibinfo {author} {\bibfnamefont {P.}~\bibnamefont {Li}}, \bibinfo {author} {\bibfnamefont {N.~W.}\ \bibnamefont {Phillips}}, \bibinfo {author} {\bibfnamefont {S.}~\bibnamefont {Leake}}, \bibinfo {author} {\bibfnamefont {M.}~\bibnamefont {Allain}}, \bibinfo {author} {\bibfnamefont {F.}~\bibnamefont {Hofmann}},\ and\ \bibinfo {author} {\bibfnamefont {V.}~\bibnamefont {Chamard}},\ }\bibfield  {title} {\bibinfo {title} {{Revealing nano-scale lattice distortions in implanted material with 3D Bragg ptychography}},\ }\href@noop {} {\bibfield  {journal} {\bibinfo  {journal} {Nature Communications}\ }\textbf {\bibinfo {volume} {12}},\ \bibinfo {pages} {7059} (\bibinfo {year} {2021})}\BibitemShut {NoStop}%
\bibitem [{\citenamefont {Godard}\ \emph {et~al.}(2011)\citenamefont {Godard}, \citenamefont {Carbone}, \citenamefont {Allain}, \citenamefont {Mastropietro}, \citenamefont {Chen}, \citenamefont {Capello}, \citenamefont {Diaz}, \citenamefont {Metzger}, \citenamefont {Stangl},\ and\ \citenamefont {Chamard}}]{godard_nat_11}%
  \BibitemOpen
  \bibfield  {author} {\bibinfo {author} {\bibfnamefont {P.}~\bibnamefont {Godard}}, \bibinfo {author} {\bibfnamefont {G.}~\bibnamefont {Carbone}}, \bibinfo {author} {\bibfnamefont {M.}~\bibnamefont {Allain}}, \bibinfo {author} {\bibfnamefont {F.}~\bibnamefont {Mastropietro}}, \bibinfo {author} {\bibfnamefont {G.}~\bibnamefont {Chen}}, \bibinfo {author} {\bibfnamefont {L.}~\bibnamefont {Capello}}, \bibinfo {author} {\bibfnamefont {A.}~\bibnamefont {Diaz}}, \bibinfo {author} {\bibfnamefont {T.~H.}\ \bibnamefont {Metzger}}, \bibinfo {author} {\bibfnamefont {J.}~\bibnamefont {Stangl}},\ and\ \bibinfo {author} {\bibfnamefont {V.}~\bibnamefont {Chamard}},\ }\bibfield  {title} {\bibinfo {title} {{Three-dimensional X-ray Bragg ptychography: high-resolution imaging of extended crystalline nanostructures}},\ }\href@noop {} {\bibfield  {journal} {\bibinfo  {journal} {Nature Communications}\ }\textbf {\bibinfo {volume} {2}},\ \bibinfo {pages} {568} (\bibinfo {year} {2011})}\BibitemShut {NoStop}%
\bibitem [{\citenamefont {Hruszkewycz}\ \emph {et~al.}(2017)\citenamefont {Hruszkewycz}, \citenamefont {Allain}, \citenamefont {Holt}, \citenamefont {Murray}, \citenamefont {Holt}, \citenamefont {Fuoss},\ and\ \citenamefont {Chamard}}]{Hru16}%
  \BibitemOpen
  \bibfield  {author} {\bibinfo {author} {\bibfnamefont {S.~O.}\ \bibnamefont {Hruszkewycz}}, \bibinfo {author} {\bibfnamefont {M.}~\bibnamefont {Allain}}, \bibinfo {author} {\bibfnamefont {M.~V.}\ \bibnamefont {Holt}}, \bibinfo {author} {\bibfnamefont {C.~E.}\ \bibnamefont {Murray}}, \bibinfo {author} {\bibfnamefont {J.~R.}\ \bibnamefont {Holt}}, \bibinfo {author} {\bibfnamefont {P.~H.}\ \bibnamefont {Fuoss}},\ and\ \bibinfo {author} {\bibfnamefont {V.}~\bibnamefont {Chamard}},\ }\bibfield  {title} {\bibinfo {title} {{High resolution three dimensional structural microscopy by single angle Bragg ptychography}},\ }\href@noop {} {\bibfield  {journal} {\bibinfo  {journal} {Nature Materials}\ }\textbf {\bibinfo {volume} {16}},\ \bibinfo {pages} {244} (\bibinfo {year} {2017})}\BibitemShut {NoStop}%
\bibitem [{\citenamefont {Guizar-Sicairos}\ and\ \citenamefont {Chamard}()}]{guizar_sicairos_2026}%
  \BibitemOpen
  \bibfield  {author} {\bibinfo {author} {\bibfnamefont {M.}~\bibnamefont {Guizar-Sicairos}}\ and\ \bibinfo {author} {\bibfnamefont {V.}~\bibnamefont {Chamard}},\ }\bibfield  {title} {\bibinfo {title} {X-ray ptychography},\ }\href@noop {} {\bibinfo  {journal} {Optica in preparation}\ }\BibitemShut {NoStop}%
\bibitem [{See Supplemental Material at [URL will be inserted by publisher] for sections S1 to S4()}]{Supplemental_Materials}%
  \BibitemOpen
\bibfield  {journal} {  }See Supplemental Material at [URL will be inserted by publisher] for sections S1 to S4,\ \href@noop {} {}\BibitemShut {NoStop}%
\bibitem [{\citenamefont {Pateras}\ \emph {et~al.}(2015)\citenamefont {Pateras}, \citenamefont {Allain}, \citenamefont {Godard}, \citenamefont {Largeau}, \citenamefont {Patriarche}, \citenamefont {Talneau}, \citenamefont {Pantzas}, \citenamefont {Burghammer}, \citenamefont {Minkevich},\ and\ \citenamefont {Chamard}}]{Pateras15}%
  \BibitemOpen
  \bibfield  {author} {\bibinfo {author} {\bibfnamefont {A.~I.}\ \bibnamefont {Pateras}}, \bibinfo {author} {\bibfnamefont {M.}~\bibnamefont {Allain}}, \bibinfo {author} {\bibfnamefont {P.}~\bibnamefont {Godard}}, \bibinfo {author} {\bibfnamefont {L.}~\bibnamefont {Largeau}}, \bibinfo {author} {\bibfnamefont {G.}~\bibnamefont {Patriarche}}, \bibinfo {author} {\bibfnamefont {A.}~\bibnamefont {Talneau}}, \bibinfo {author} {\bibfnamefont {K.}~\bibnamefont {Pantzas}}, \bibinfo {author} {\bibfnamefont {M.}~\bibnamefont {Burghammer}}, \bibinfo {author} {\bibfnamefont {A.~A.}\ \bibnamefont {Minkevich}},\ and\ \bibinfo {author} {\bibfnamefont {V.}~\bibnamefont {Chamard}},\ }\bibfield  {title} {\bibinfo {title} {Nondestructive three-dimensional imaging of crystal strain and rotations in an extended bonded semiconductor heterostructure},\ }\href@noop {} {\bibfield  {journal} {\bibinfo  {journal} {Physical Review B}\ }\textbf {\bibinfo {volume} {92}},\ \bibinfo {pages} {205305} (\bibinfo {year} {2015})}\BibitemShut
  {NoStop}%
\bibitem [{\citenamefont {Calvo-Almaz\'an}\ \emph {et~al.}(2024)\citenamefont {Calvo-Almaz\'an}, \citenamefont {Chamard}, \citenamefont {Gr\"unewald},\ and\ \citenamefont {Allain}}]{Calvo24}%
  \BibitemOpen
  \bibfield  {author} {\bibinfo {author} {\bibfnamefont {I.}~\bibnamefont {Calvo-Almaz\'an}}, \bibinfo {author} {\bibfnamefont {V.}~\bibnamefont {Chamard}}, \bibinfo {author} {\bibfnamefont {T.~A.}\ \bibnamefont {Gr\"unewald}},\ and\ \bibinfo {author} {\bibfnamefont {M.}~\bibnamefont {Allain}},\ }\bibfield  {title} {\bibinfo {title} {{Inhomogeneous probes for Bragg coherent diffraction imaging: Toward the imaging of dynamic and distorted crystals}},\ }\href {https://doi.org/10.1103/PhysRevB.110.134117} {\bibfield  {journal} {\bibinfo  {journal} {Phys. Rev. B}\ }\textbf {\bibinfo {volume} {110}},\ \bibinfo {pages} {134117} (\bibinfo {year} {2024})}\BibitemShut {NoStop}%
\bibitem [{\citenamefont {Newton}(2020)}]{Newton_multiple_2020}%
  \BibitemOpen
  \bibfield  {author} {\bibinfo {author} {\bibfnamefont {M.~C.}\ \bibnamefont {Newton}},\ }\bibfield  {title} {\bibinfo {title} {Concurrent phase retrieval for imaging strain in nanocrystals},\ }\href {https://doi.org/10.1103/PhysRevB.102.014104} {\bibfield  {journal} {\bibinfo  {journal} {Phys. Rev. B}\ }\textbf {\bibinfo {volume} {102}},\ \bibinfo {pages} {014104} (\bibinfo {year} {2020})}\BibitemShut {NoStop}%
\bibitem [{\citenamefont {Masto}\ \emph {et~al.}()\citenamefont {Masto}, \citenamefont {Favre-Nicolin}, \citenamefont {Leake}, \citenamefont {Sch¨ulli}, \citenamefont {Richard}, \citenamefont {Atlan},\ and\ \citenamefont {Bellec4}}]{Masto_deep_learning_2026}%
  \BibitemOpen
  \bibfield  {author} {\bibinfo {author} {\bibfnamefont {M.}~\bibnamefont {Masto}}, \bibinfo {author} {\bibfnamefont {V.}~\bibnamefont {Favre-Nicolin}}, \bibinfo {author} {\bibfnamefont {S.}~\bibnamefont {Leake}}, \bibinfo {author} {\bibfnamefont {T.}~\bibnamefont {Sch¨ulli}}, \bibinfo {author} {\bibfnamefont {M.-I.}\ \bibnamefont {Richard}}, \bibinfo {author} {\bibfnamefont {C.}~\bibnamefont {Atlan}},\ and\ \bibinfo {author} {\bibfnamefont {E.}~\bibnamefont {Bellec4}},\ }\bibfield  {title} {\bibinfo {title} {Phase retrieval of highly strained bragg coherent diffraction patterns using supervised convolutional neural network},\ }\bibinfo {note} {under review}\BibitemShut {NoStop}%
\bibitem [{\citenamefont {Sun}\ and\ \citenamefont {Singer}(2024)}]{Sun_review_2024}%
  \BibitemOpen
  \bibfield  {author} {\bibinfo {author} {\bibfnamefont {Y.}~\bibnamefont {Sun}}\ and\ \bibinfo {author} {\bibfnamefont {A.}~\bibnamefont {Singer}},\ }\bibfield  {title} {\bibinfo {title} {Bragg coherent diffractive imaging for defects analysis: Principles, applications, and challenges},\ }\href {https://doi.org/10.1063/5.0219030} {\bibfield  {journal} {\bibinfo  {journal} {Chemical Physics Reviews}\ }\textbf {\bibinfo {volume} {5}},\ \bibinfo {pages} {031310} (\bibinfo {year} {2024})}\BibitemShut {NoStop}%
\bibitem [{\citenamefont {Li}\ \emph {et~al.}(2022)\citenamefont {Li}, \citenamefont {Allain}, \citenamefont {Gruenewald}, \citenamefont {Rommel}, \citenamefont {Campos}, \citenamefont {Carbone},\ and\ \citenamefont {Chamard}}]{Li22}%
  \BibitemOpen
  \bibfield  {author} {\bibinfo {author} {\bibfnamefont {P.}~\bibnamefont {Li}}, \bibinfo {author} {\bibfnamefont {M.}~\bibnamefont {Allain}}, \bibinfo {author} {\bibfnamefont {T.~A.}\ \bibnamefont {Gruenewald}}, \bibinfo {author} {\bibfnamefont {M.}~\bibnamefont {Rommel}}, \bibinfo {author} {\bibfnamefont {A.}~\bibnamefont {Campos}}, \bibinfo {author} {\bibfnamefont {D.}~\bibnamefont {Carbone}},\ and\ \bibinfo {author} {\bibfnamefont {V.}~\bibnamefont {Chamard}},\ }\bibfield  {title} {\bibinfo {title} {4th generation synchrotron source boosts crystalline imaging at the nanoscale},\ }\href@noop {} {\bibfield  {journal} {\bibinfo  {journal} {Light: Science et applications}\ }\textbf {\bibinfo {volume} {11}},\ \bibinfo {pages} {73} (\bibinfo {year} {2022})}\BibitemShut {NoStop}%
\bibitem [{\citenamefont {Hruszkewycz}\ \emph {et~al.}(2013)\citenamefont {Hruszkewycz}, \citenamefont {Highland}, \citenamefont {Holt}, \citenamefont {Kim}, \citenamefont {Folkman}, \citenamefont {Thompson}, \citenamefont {Tripathi}, \citenamefont {Stephenson}, \citenamefont {Hong},\ and\ \citenamefont {Fuoss}}]{Hruszkewycz_ferro_2013}%
  \BibitemOpen
  \bibfield  {author} {\bibinfo {author} {\bibfnamefont {S.~O.}\ \bibnamefont {Hruszkewycz}}, \bibinfo {author} {\bibfnamefont {M.~J.}\ \bibnamefont {Highland}}, \bibinfo {author} {\bibfnamefont {M.~V.}\ \bibnamefont {Holt}}, \bibinfo {author} {\bibfnamefont {D.}~\bibnamefont {Kim}}, \bibinfo {author} {\bibfnamefont {C.~M.}\ \bibnamefont {Folkman}}, \bibinfo {author} {\bibfnamefont {C.}~\bibnamefont {Thompson}}, \bibinfo {author} {\bibfnamefont {A.}~\bibnamefont {Tripathi}}, \bibinfo {author} {\bibfnamefont {G.~B.}\ \bibnamefont {Stephenson}}, \bibinfo {author} {\bibfnamefont {S.}~\bibnamefont {Hong}},\ and\ \bibinfo {author} {\bibfnamefont {P.~H.}\ \bibnamefont {Fuoss}},\ }\bibfield  {title} {\bibinfo {title} {{Imaging Local Polarization in Ferroelectric Thin Films by Coherent X-Ray Bragg Projection Ptychography}},\ }\href {https://doi.org/10.1103/PhysRevLett.110.177601} {\bibfield  {journal} {\bibinfo  {journal} {Phys. Rev. Lett.}\ }\textbf {\bibinfo {volume} {110}},\ \bibinfo {pages} {177601} (\bibinfo
  {year} {2013})}\BibitemShut {NoStop}%
\bibitem [{\citenamefont {Thibault}\ \emph {et~al.}(2008)\citenamefont {Thibault}, \citenamefont {Dierolf}, \citenamefont {Menzel}, \citenamefont {Bunk}, \citenamefont {David},\ and\ \citenamefont {Pfeiffer}}]{thibault08}%
  \BibitemOpen
  \bibfield  {author} {\bibinfo {author} {\bibfnamefont {P.}~\bibnamefont {Thibault}}, \bibinfo {author} {\bibfnamefont {M.}~\bibnamefont {Dierolf}}, \bibinfo {author} {\bibfnamefont {A.}~\bibnamefont {Menzel}}, \bibinfo {author} {\bibfnamefont {O.}~\bibnamefont {Bunk}}, \bibinfo {author} {\bibfnamefont {C.}~\bibnamefont {David}},\ and\ \bibinfo {author} {\bibfnamefont {F.}~\bibnamefont {Pfeiffer}},\ }\bibfield  {title} {\bibinfo {title} {High-resolution scanning x-ray diffraction microscopy},\ }\href@noop {} {\bibfield  {journal} {\bibinfo  {journal} {Science}\ }\textbf {\bibinfo {volume} {321}},\ \bibinfo {pages} {379} (\bibinfo {year} {2008})}\BibitemShut {NoStop}%
\bibitem [{\citenamefont {Li}\ \emph {et~al.}(2020)\citenamefont {Li}, \citenamefont {Maddali}, \citenamefont {Pateras}, \citenamefont {Calvo-Almazan}, \citenamefont {Hruszkewycz}, \citenamefont {Cha}, \citenamefont {Chamard},\ and\ \citenamefont {Allain}}]{Li2020_formalism}%
  \BibitemOpen
  \bibfield  {author} {\bibinfo {author} {\bibfnamefont {P.}~\bibnamefont {Li}}, \bibinfo {author} {\bibfnamefont {S.}~\bibnamefont {Maddali}}, \bibinfo {author} {\bibfnamefont {A.}~\bibnamefont {Pateras}}, \bibinfo {author} {\bibfnamefont {I.}~\bibnamefont {Calvo-Almazan}}, \bibinfo {author} {\bibfnamefont {S.}~\bibnamefont {Hruszkewycz}}, \bibinfo {author} {\bibfnamefont {W.}~\bibnamefont {Cha}}, \bibinfo {author} {\bibfnamefont {V.}~\bibnamefont {Chamard}},\ and\ \bibinfo {author} {\bibfnamefont {M.}~\bibnamefont {Allain}},\ }\bibfield  {title} {\bibinfo {title} {{General approaches for shear-correcting coordinate transformations in Bragg coherent diffraction imaging. Part II}},\ }\href {https://doi.org/10.1107/S1600576720001375} {\bibfield  {journal} {\bibinfo  {journal} {Journal of Applied Crystallography}\ }\textbf {\bibinfo {volume} {53}},\ \bibinfo {pages} {404} (\bibinfo {year} {2020})}\BibitemShut {NoStop}%
\bibitem [{\citenamefont {Chen}\ \emph {et~al.}(2007)\citenamefont {Chen}, \citenamefont {Miao}, \citenamefont {Wang},\ and\ \citenamefont {Lee}}]{Chen_2007_gHIO}%
  \BibitemOpen
  \bibfield  {author} {\bibinfo {author} {\bibfnamefont {C.-C.}\ \bibnamefont {Chen}}, \bibinfo {author} {\bibfnamefont {J.}~\bibnamefont {Miao}}, \bibinfo {author} {\bibfnamefont {C.~W.}\ \bibnamefont {Wang}},\ and\ \bibinfo {author} {\bibfnamefont {T.~K.}\ \bibnamefont {Lee}},\ }\bibfield  {title} {\bibinfo {title} {Application of optimization technique to noncrystalline x-ray diffraction microscopy: Guided hybrid input-output method},\ }\href {https://doi.org/10.1103/PhysRevB.76.064113} {\bibfield  {journal} {\bibinfo  {journal} {Phys. Rev. B}\ }\textbf {\bibinfo {volume} {76}},\ \bibinfo {pages} {064113} (\bibinfo {year} {2007})}\BibitemShut {NoStop}%
\bibitem [{\citenamefont {Fienup}(1982)}]{fienup82}%
  \BibitemOpen
  \bibfield  {author} {\bibinfo {author} {\bibfnamefont {J.~R.}\ \bibnamefont {Fienup}},\ }\bibfield  {title} {\bibinfo {title} {Phase retrieval algorithms: a comparison},\ }\href@noop {} {\bibfield  {journal} {\bibinfo  {journal} {Appl. Opt.}\ }\textbf {\bibinfo {volume} {21}},\ \bibinfo {pages} {2758} (\bibinfo {year} {1982})}\BibitemShut {NoStop}%
\bibitem [{\citenamefont {Marchesini}\ \emph {et~al.}(2003)\citenamefont {Marchesini}, \citenamefont {He}, \citenamefont {Chapman}, \citenamefont {Hau-Riege}, \citenamefont {Noy}, \citenamefont {Howells}, \citenamefont {Weierstall},\ and\ \citenamefont {Spence}}]{marchesini_shrinkwrap_2003}%
  \BibitemOpen
  \bibfield  {author} {\bibinfo {author} {\bibfnamefont {S.}~\bibnamefont {Marchesini}}, \bibinfo {author} {\bibfnamefont {H.}~\bibnamefont {He}}, \bibinfo {author} {\bibfnamefont {H.~N.}\ \bibnamefont {Chapman}}, \bibinfo {author} {\bibfnamefont {S.~P.}\ \bibnamefont {Hau-Riege}}, \bibinfo {author} {\bibfnamefont {A.}~\bibnamefont {Noy}}, \bibinfo {author} {\bibfnamefont {M.~R.}\ \bibnamefont {Howells}}, \bibinfo {author} {\bibfnamefont {U.}~\bibnamefont {Weierstall}},\ and\ \bibinfo {author} {\bibfnamefont {J.~C.~H.}\ \bibnamefont {Spence}},\ }\bibfield  {title} {\bibinfo {title} {X-ray image reconstruction from a diffraction pattern alone},\ }\href {https://doi.org/10.1103/PhysRevB.68.140101} {\bibfield  {journal} {\bibinfo  {journal} {Phys. Rev. B}\ }\textbf {\bibinfo {volume} {68}},\ \bibinfo {pages} {140101} (\bibinfo {year} {2003})}\BibitemShut {NoStop}%
\end{thebibliography}
%

\newpage
\appendix
\section{Appendix A: Sample preparation}
\label{Appendix_sample_prep}
The production of crystalline isolated particles involved the dewetting of a 20~nm thick gold layer, thermally evaporated onto a silicon wafer covered with a 2~nm thick titanium adhesion layer, before the sample was annealed (1273~K, 10~h in air) to form the micro-crystals. 

\section{Appendix B: Measurement geometry and data acquisition}
The incident X-ray photon energy was set to 11.8~keV using a Si-111 double crystal monochromator and further focused using a Fresnel zone plate, with outmost zone width of 200~nm and diameter of 400~$\mu$m. It produced a focal spot of about 250~nm. The beam was fully characterized with standard forward ptychography prior mounting the Au sample \cite{thibault08}. At the 3DBP sample position, a few millimeters downstream the focal plane, the beam was a 2~$\mu$m diameter circular divergent wavefield (see Fig.~\ref{fig:1}(c) and Supplemental Materials, section S1 \cite{Supplemental_Materials}). For BCDI, the lens aperture was reduced with a set of slits ($20 \times 20~\mu m^2$), placed upstream and off-centered with respect to the optical axis (Fig.~\ref{fig:1}(b)), resulting in a $2\times 2~\mu m^2$ focal sport with rather constant amplitude and phase. For both BCDI and 3DBP data acquisition, the vertically mounted sample (with its substrate parallel to the incident beam) was rotated to the (111) Bragg angle (12.9$^\circ$) with respect to the beam direction. The area detector (ExcaliburRX with 55 $\mu m$ pixel size and $2048 \times 1536$ pixels) was positioned at twice the Bragg angle (25.8$^\circ$) and 2.8~m away from the sample. For the BCDI measurements, an angular rocking scan with a step size of 0.005$^\circ$ was carried out, rotating the sample about the vertical axis to cover an angular range of 0.4$^\circ$. At each angular position, the diffraction pattern was recorded with 10~s exposure time. For 3DBP, an additional raster grid scan of $11 \times 11$ points with 200~nm step on the sample was performed at each angle, with an exposure time per point set to 0.1~s. For the weakly distorted crystal, only the central 60 angles were used for both the BCDI and 3DBP reconstructions to exclude an alien peak appeared at the end of the rocking curve. The 3DBP data sets of the weakly and highly  distorted crystal contain $1.7 \times 10^7$ and $2.1 \times 10^7$ photons, respectively, while $2.5 \times 10^7$ and $2.2 \times 10^7$ photons were collected in their respective BCDI data sets, taking into account of the excluding angles for the case of weakly distorted crystal.  

\section{Appendix C: Inversion procedure}
\label{Appendix_inversion}
For 3DBP, 400 iterations were run using the ptychography iterative engine (\textit{i.e.}, ordered subset with Gaussian noise model) \cite{li21}. The reconstruction was performed in the orthogonal frame defined by the detector plane and the exit beam direction. This direct mapping of the real and reciprocal space frames was enabled via a modified 3D Fourier transform recently developed \cite{Li2020_formalism}. The inversion retrieves both the probe and the object functions simultaneously. For the probe retrieval, an invariance constraint was applied \cite{li21} along the beam propagation direction since the depth of focus for the zone plate used here was about 500~$\mu$m, much larger than the intersected volume with the object. For the object retrieval, a thickness-support regularization was applied to avoid the inherent under-determination along the beam propagation direction \cite{li21}. Meanwhile, an amplitude threshold was also implemented for the object reconstruction: during the first 150 iterations, the relative object amplitudes lower than 10\% of the maximum were set to zero and higher than the threshold set to 1 every 5 iterations. An incoherent background contribution is jointly fitted during the inversion to account for incoherent photon counts, supposedly induced by air scattering.

The BCDI data sets were inverted with phase retrieval code based from previously published work \cite{Clark_dissolution_2015}. A guided phase-retrieval approach \cite{Chen_2007_gHIO} with 40 random starts and four generations was used, with a best solution selection based on a sharpness metric. This was previously shown to yield the most truthful reconstructions for strained samples \cite{Clark_dissolution_2015}. For each generation, 620 phase-retrieval iterations were performed consisting of a pattern of 20 iterations of error reduction (ER) and 180 iterations hybrid input–output (HIO) \cite{fienup82} repeated three times, followed by a final 20 iterations of ER. The support was updated every iteration using the Shrinkwrap algorithm \cite{marchesini_shrinkwrap_2003} with a threshold of 0.1 and a Gaussian kernel with sigma of 1 pixel.

For all reconstructed results shown, the amplitude of the retrieved object was normalized such that its maximum value equals 1. For both the 3DBP and BCDI reconstructions, the measured diffraction patterns were binned by $2 \times 2$.

\end{document}